\documentclass[11pt,a4paper]{article}
\usepackage{amsmath, amsfonts, amssymb}
\usepackage{a4wide}
\usepackage[left=0.8in,bottom=1in,top=1in,right=0.8in]{geometry}
\usepackage{parskip}
\usepackage{enumitem}
\usepackage{xcolor}

\usepackage{hyperref}
\usepackage{booktabs}
\usepackage{caption}
\usepackage{siunitx}
\usepackage{tabularx}
\usepackage{comment}
\usepackage{graphicx}
\usepackage{adjustbox}
\usepackage{multirow,tensor}
\usepackage{amsthm}
\usepackage{bm, makecell}
\usepackage{lscape}
\usepackage{rotating}
\usepackage{floatrow}
\usepackage[noadjust]{cite}

\def\qed{\hfill {$\square$}\goodbreak \medskip}

\newtheorem{theorem}{Theorem}[section]
\newtheorem{lemma}[theorem]{Lemma}
\newtheorem{corollary}[theorem]{Corollary}

\newtheorem{definition}[theorem]{Definition}
\newtheorem{example}[theorem]{Example}
\newtheorem{remark}[theorem]{Remark}

\numberwithin{equation}{section}

\newcommand{\Tr}{\textnormal{Tr}}
\newcommand{\Supp}{\textnormal{Supp}}
\DeclareMathOperator{\wt}{wt}

\usepackage{tikz,xcolor,hyperref}
\usepackage{mathdots}

\definecolor{lime}{HTML}{A6CE39}
\DeclareRobustCommand{\orcidicon}{%
	\begin{tikzpicture}
		\draw[lime, fill=lime] (0,0) 
		circle [radius=0.16] 
		node[white] {{\fontfamily{qag}\selectfont \tiny ID}};
		\draw[white, fill=white] (-0.0625,0.095) 
		circle [radius=0.007];
	\end{tikzpicture}
	\hspace{-2mm}
}

\foreach \x in {A, ..., Z}{%
	\expandafter\xdef\csname orcid\x\endcsname{\noexpand\href{https://orcid.org/\csname orcidauthor\x\endcsname}{\noexpand\orcidicon}}
}

\begin{document}
	\date{}
		\title{Optimal binary codes from $\mathcal{C}_{D}$-codes over a non-chain ring}

		\author{{\bf Ankit Yadav\footnote{email: {\tt ankityadav10102000@gmail.com}}\orcidA{},\; 
        Ritumoni Sarma\footnote{email: {\tt ritumoni407@gmail.com}}\orcidB{} and \bf Anuj Kumar Bhagat\footnote{email: {\tt anujkumarbhagat632@gmail.com}}\orcidC{}} \\ Department of Mathematics\\ Indian Institute of Technology Delhi\\Hauz Khas, New Delhi-110016, India \medskip \\
  }
  
\maketitle

\begin{abstract}
In \cite{shi2022few-weight}, Shi and Li studied $\mathcal{C}_D$-codes over the ring $\mathcal{R}:=\mathbb{F}_2[x,y]/\langle x^2, y^2, xy-yx\rangle$ and their binary Gray images, where $D$ is derived using certain simplicial complexes. We study the subfield codes $\mathcal{C}_{D}^{(2)}$ of $\mathcal{C}_{D}$-codes over $\mathcal{R},$ where $D$ is as in \cite{shi2022few-weight} and more. We find the Hamming weight distribution and the parameters of $\mathcal{C}_D^{(2)}$ for various $D$, and identify several infinite families of codes that are distance-optimal. Besides, we provide sufficient conditions under which these codes are minimal and self-orthogonal. Two families of strongly regular graphs are obtained as an application of the constructed two-weight codes. 

\medskip

\medskip
			

\end{abstract}

\section{Introduction}\label{Section1}
A linear code with parameters $[n, k, d]$ is said to be distance-optimal if there are no linear codes with parameters $[n, k, d+1]$ over the same finite field. For fixed parameters $n$ and $k$, distance-optimal codes achieve the largest possible error-detection and error-correction capabilities. Indeed, a main purpose of coding theory is to give an explicit construction of distance-optimal codes. In this direction, there are several methods in the literature. For instance, Gray maps are helpful for constructing codes over a finite field from codes defined over a finite ring. Any distance-preserving map from a ring to a field is called a Gray map. Hammons et al. \cite{Hammons1994} were the first to apply this idea to construct optimal families of nonlinear codes over the binary field. Since then, many researchers have obtained various optimal codes using Gray maps, see, for instance, \cite{Wu2024binaryimage,Yadav2025ACD,habibul2022lcd,Shi2022Polycyclic,shi2017constacyclic}. Another important method to construct optimal linear codes involves the use of simplicial complexes. Chang and Hyun \cite{Chang_Hyun2018simplicial} were the first to introduce this approach, to construct optimal linear codes and to determine their weight distributions. Subsequently, many researchers have applied simplicial complexes to construct a wide range of optimal codes over different finite fields (see \cite{anuj2025subfield,anuj2025binary,heng2020two,wu2020optimal,wu2022quaternary,Ding2019subfield,shi2017two,shi2021two,li2020new}).

Given a multiset $D$ over a finite commutative $\mathbb{F}_q$-algebra, we can construct a linear code, denoted by $\mathcal{C}_D$, having length $|D|$. The weight distribution of $\mathcal{C}_D$ can be computed efficiently if $D$ is constructed using a simplicial complex, and there is substantial evidence that $\mathcal{C}_D$ possesses good parameters (see \cite{Liu2024,vidya2024nonunital,Mondal2024mixed_alphabet}). For these reasons, simplicial complexes have attracted significant interest in algebraic coding theory. For determining the error-correction capability, the weight distribution of a linear code plays a crucial role. Therefore, calculating the weight distribution of a linear code is of fundamental importance.

A non-zero vector $v$ is minimal if for any other vector $u$ which is not a nonzero scalar multiple of $v$, the support of $u$ is not a subset of the support of $v$. Determining the full set of minimal codewords, known as the covering problem, is generally a hard problem \cite{Huffman2021Concise}. A linear code whose every nonzero codeword is minimal is considered important; such linear codes are called minimal. In fact, in secret-sharing schemes, such codes are extensively used (see \cite{shamir1979secret, yuan2006secret}), where they define the entire access structure. They are also used in secure two-party computation protocols \cite{chabanne2014twoparty}, where they ensure the privacy of the parties involved. Moreover, the minimum distance decoding method \cite{Ashikhmin1998} can be used efficiently to decode the minimal linear codes. This makes the study and construction of minimal linear codes an active area of research (see \cite{Vidya2024minimal,Hyun2020optimal,maji2025minimal}).

In \cite{shi2022few-weight}, the authors investigated the Lee weight distribution of the $\mathcal{C}_D$-codes over the commutative $\mathbb{F}_2$-algebra $\mathcal{R} := \mathbb{F}_2 + u\mathbb{F}_2 + v\mathbb{F}_2 + uv\mathbb{F}_2$ with $u^2 = v^2 = 0$, by choosing the defining set $D$ based on simplicial complexes with only one maximal element. They also studied their Gray images and derived various families of binary linear codes which are both minimal and optimal. We study the subfield codes $\mathcal{C}_D^{(2)}$ of $\mathcal{C}_D$-codes defined over $\mathcal{R}$, where $D$ is derived from simplicial complexes generated by a single element. We determine the Hamming weight distribution and the parameters of $\mathcal{C}_D^{(2)}.$ We show that the binary subfield codes of $\mathcal{C}_D$ have improved parameters as compared to the binary Gray images of $\mathcal{C}_D$ as considered in \cite{shi2022few-weight}. Furthermore, we prove that for any chosen defining set $D$ derived using simplicial complexes, all constructed families of $\mathcal{C}_D^{(2)}$ are minimal and self-orthogonal under mild conditions and certain families are distance-optimal. As a concrete application, we employ the techniques in \cite{calderbank1986SRG} to construct families of strongly regular graphs from certain two-weight projective $\mathcal{C}_D^{(2)}$ codes.

The subsequent sections are arranged as follows: the forthcoming section presents notations and preliminaries. Section \ref{Section3} describes the binary subfield codes $\mathcal{C}_D^{(2)}$ of the $\mathcal{C}_D$-codes over $\mathcal{R}$. Section \ref{Section 4} presents the Hamming weight distribution of $\mathcal{C}_D^{(2)}$ for various $D$. Additionally, minimality and self-orthogonality conditions for $\mathcal{C}_D^{(2)}$ are established in Section \ref{Section 4}. Section \ref{Section 5} is the concluding section of the article.

\section{Preliminaries}\label{Section2}
We use the following notations throughout this article for a natural number $m$ and a vector $v\in\mathbb{F}_q^n$.
\begin{table}[H]
\centering
\begin{tabular}{l|l}
\hline
Notation & Description \\
\hline
$[m]$ & the set $\{1, 2, \dots, m\}$\\
$\#S$ &  the cardinality of set $S$ \\
$\mathcal{P}(S)$ & the power set of set $S$\\
$\Supp(v)$ & $\{j\in [n]\;|\; v_j\ne 0\}$\\
$\wt(v)$ & $\#\Supp (v)$\\
$\Delta_X$ & $\{ u \in \mathbb{F}_{2}^{m} \mid \Supp(u) \subseteq X\}$ for $X\subseteq[m]$\\
$\mathcal{R}$ & the $\mathbb{F}_{2}$-algebra $\mathbb{F}_{2}[x,y]/\langle x^2, y^2, xy-yx \rangle$\\
$\delta_{i, j}$ & the Kronecker delta function\\
\hline
\end{tabular}
\end{table}
Let $R$ be a finite commutative ring. \textit{Linear codes} of length $n$ are $R$-submodules of $R^n$. When $R=\mathbb{F}_{q}$, the parameters of the linear code $C$ are denoted by $[n,k,d]$, where $k$ and $d$ represent the dimension and the minimum Hamming weight among all the non-zero codewords of $C$, respectively. Let $A_{j}$ be the number of codewords of $C$ having weight $j$, then the string $(A_{0},A_{1},\ldots,A_{n})$ is said to be the \textit{Hamming weight distribution} of $C$. The code $C$  is called \textit{$t$-weight} linear code if only $t$ entries among $(A_{1},A_{2},\ldots,A_{n})$ are non-zero. The linear code $C$ is said to be \textit{projective} if $d(C^{\perp})\ge 3.$

Next, we have a well-known Griesmer bound which ensures the distance-optimality of a linear code. A linear code attaining this bound is called a \textit{Griesmer code}.
\begin{lemma}\cite{Griesmer1960bound}
    Every linear code over the field $\mathbb{F}_{q}$ with parameters $[n,k,d]$ satisfies $$n \geq \sum\limits_{i=0}^{k-1} \left\lceil \frac{d}{q^{i}} \right\rceil.$$
\end{lemma}  
\begin{theorem}\cite{Huffmann_pless} \label{self_orthogonality_theorem}
A binary linear code is self-orthogonal with respect to the Euclidean inner product if the Hamming weight of each of its codewords is a multiple of $4$.
\end{theorem}

Let $v,w \in \mathbb{F}_{q}^{n}$. Then, $w$ \textit{covers} $v$ (we write $v \preceq w$) if $\Supp(w) \supseteq \Supp(v)$. Note that $v\preceq a v,$ for any $a\in\mathbb{F}_q^{*}.$
For a linear code $C$ and a non-zero codeword 
$v$, if $v\preceq w,$ for $w\in C$ implies $w=av,$ for some $a\in\mathbb{F}_q^*,$ then we say $v$ is \textit{minimal}. If every non-zero codeword is minimal, then the code itself is called {\it minimal}.

Next, we have a lemma for the minimality of a linear code.
\begin{lemma}\cite{Ashikhmin1998}
    Let $\wt_{\min}$ and $\wt_{\max}$ denote the minimum and maximum non-zero weights of a linear code $C$ over $\mathbb{F}_{q}$, respectively. 
    If $\frac{\wt_{\min}}{\wt_{\max}} > \frac{q-1}{q}$, then $C$ is minimal.
\end{lemma}
Every element of $\mathcal{R}$ is of the form $ a+ bu + cv + duv $, for $a,b,c,d \in \mathbb{F}_{2}$ and $u = x +\langle x^2, y^2 \rangle, v = y + \langle x^2, y^2 \rangle$. For an ordered multiset $D=\{\{d_{1}<d_{2}<\ldots<d_{n}\}\}$, where $d_{j} \in \mathcal{R}^{m}$ for $1\leq j \leq n$, define
\begin{equation*}
    \mathcal{C}_{D} := \{(\mathbf{z}\cdot d_{1},\mathbf{z}\cdot d_{2},\ldots,\mathbf{z}\cdot d_{n}) \mid \mathbf{z} \in \mathcal{R}^{m}\},
\end{equation*}
where $\cdot$ denotes the Euclidean inner product on $\mathcal{R}^{m}$. Then, $\mathcal{C}_{D}$ is a linear code over $\mathcal{R}$ of length $n$ and $D$ is said to be the \textit{defining set} of $\mathcal{C}_{D}$.
\subsection{Simplicial complex}
The map $\psi : \mathbb{F}_{2}^{m} \to \mathcal{P}([m])$ given by $v\mapsto \Supp(v)$ is a bijective map. Thus, whenever convenient, we denote $\Supp(v)$ by simply $v$.
\begin{definition}
    A \textit{simplicial complex} $\Delta$ is a subset of $ \mathbb{F}_{2}^{m}$ that satisfies the condition that $v\in \Delta \implies w\in \Delta$ for all $w\preceq v$. An element $v \in \Delta$ is said to be \textit{maximal} if for no $w\in \Delta\setminus\{v\}$, we have $v\preceq w$.
\end{definition}
The smallest simplicial complex that contains $X\subseteq [m]$ (denoted by $\Delta_{X}$) is called the simplicial complex generated by $X$. By definition, $|\Delta_{X}| = 2^{|X|}.$

For $Q\subseteq \mathbb{F}_{2}^{m}$, define 
\begin{equation}
    \mathcal{H}_{Q}(y_{1},y_{2},\ldots,y_{m}) := \sum\limits_{v\in Q} \prod\limits_{j=1}^{m} y_{j}^{v_{j}} \in \mathbb{Z}[y_{1},y_{2},\ldots,y_{m}].
\end{equation}
\begin{lemma}\cite{Chang_Hyun2018simplicial}
     Suppose the set $\mathcal{F}$ contains all maximal elements of the simplicial complex $\Delta\subseteq \mathbb{F}_{2}^{m}$. Then $$\mathcal{H}_{Q}(y_{1},y_{2},\ldots,y_{m}) = \sum\limits_{\emptyset \neq S \subseteq \mathcal{F}} (-1)^{|S|+1} \prod\limits_{j\in \cap S} (1+y_{j}),$$
    where $\bigcap S $ means $ \underset{F\in S}{\bigcap}\Supp(S).$ For $Q=\Delta$, we have $$|\Delta| = \sum\limits_{\emptyset \neq S \subseteq \mathcal{F}} (-1)^{|S|+1}2^{|\cap S|}.$$
\end{lemma}

\subsection{Strongly regular graphs}

A  graph $G=(V, E)$ is called a strongly regular graph (in short, SRG) with parameters $(N_1, K_1, \lambda, \mu)$ if $|V|=N_1,$ degree of each vertex is $K_1,$ any two pairs of adjacent vertices have equal number $(=\lambda)$ of common neighbours and any two pairs of non-adjacent vertices have equal number ($=\mu$) of common neighbours.
The correspondence between projective two-weight linear codes and strong regular graphs is shown in \cite{calderbank1986SRG} as follows:
\begin{theorem}\cite{calderbank1986SRG}\label{SRGlemma}
    Suppose $C$ denotes a projective linear code of length $n$ and dimension $k$ over $\mathbb{F}_{q}$ such that $C$ is a two-weight code with non-zero weights $w_{1}$ and $w_{2}$. Then the strongly regular graph corresponding to $C$ has parameters $(N_{1},K_{1},\lambda,\mu)$ given by: 
    \begin{eqnarray*}
        N_1 &=& q^{k},\\
        K_1 &=& n(q-1),\\
        \lambda &=& K_1^{2}+3K_{1}-q(w_{1}+w_{2})-K_{1}q(w_{1}+w_{2})+q^{2}w_{1}w_2, \\
        \mu &=& \frac{q^{2}w_{1}w_{2}}{q^{k}}.
    \end{eqnarray*}
\end{theorem}
\section{Construction of Subfield Codes over $\mathcal{R}$} \label{Section3}
\begin{definition}\cite{generalized}
   Let $R$ be a subring of the ring $S$ (not necessarily commutative) such that $R$ and $S$ have the same unity. Then a surjective homomorphism of left $R$-modules $\Tr_R^S: S\to R$ is called an \textit{$R$-valued trace} of $S$ if zero is the only ideal of $S$ contained in ker$(\Tr_R^S)$.
\end{definition}
Let $R$ be a finite commutative $\mathbb{F}_{q}$-algebra with an $\mathbb{F}_{q}$-valued trace, denoted by $\tau$, and let $B$ be an $\mathbb{F}_{q}$-basis of $R$. Assume that $C$ is a linear code of length $n$ over $R$ and $G$ is a generator matrix of $C$. 
 If we replace each entry of the matrix $G$ by its column representation with respect to the basis $\mathcal{B}$, the resulting matrix generates a code known as the \textit{subfield code}, denoted by $C^{(q)}$.
 
 The $\mathbb{F}_q$-valued trace helps in constructing the subfield code.
\begin{theorem}\cite{Bhagat2024Trace} \label{trace_matrix}
Suppose $C$ denotes a linear code over $R$ and $n$ denotes its length. Suppose $G=(\mathbf{g}_{ij})_{k\times n}$ is a generator matrix of $C$, $\mathcal{B} = \{\boldsymbol{\alpha}_{1},\boldsymbol{\alpha}_{2},\ldots,\boldsymbol{\alpha}_{m}\} $ is an $\mathbb{F}_{q}$-basis of $R$ and $\tau:R \to \mathbb{F}_q$ is an $\mathbb{F}_q$-valued trace of $R$. Then $C^{(q)}$ is generated by  
\begin{center}
    $G^{(q)} =  
\left[
\begin{tabular}{c}
     $G_{1}^{(q)}$  \\
      $G_{2}^{(q)}$ \\
      $\vdots$ \\
      $G_{k}^{(q)}$
\end{tabular}
\right],$
\end{center}
where 
\begin{center}
    $G_{j}^{(q)} = \left[ 
    \begin{tabular}{c c c c}
         $\tau(\mathbf{g}_{j1}\boldsymbol{\alpha}_{1})$& $\tau(\mathbf{g}_{j2}\boldsymbol{\alpha}_{1})$ & $\cdots$ & $\tau(\mathbf{g}_{jn}\boldsymbol{\alpha}_{1})$ \\
          $\tau(\mathbf{g}_{j1}\boldsymbol{\alpha}_{2})$& $\tau(\mathbf{g}_{j2}\boldsymbol{\alpha}_{2})$ & $\cdots$ & $\tau(\mathbf{g}_{jn}\boldsymbol{\alpha}_{2})$ \\
          $\vdots$ & $\vdots$ & $\ddots$ & $\vdots$ \\
          $\tau(\mathbf{g}_{j1}\boldsymbol{\alpha}_{m})$& $\tau(\mathbf{g}_{j2}\boldsymbol{\alpha}_{m})$ & $\cdots$ & $\tau(\mathbf{g}_{jn}\boldsymbol{\alpha}_{m})$ \\
    \end{tabular}
    \right]$
\end{center}
for all $j\in [k]$.
\end{theorem}
It is easy to see that the map $\tau : \mathcal{R} \to \mathbb{F}_{2}$ defined by $a+bu+cv+duv \mapsto a+b+c+d$ is an $\mathbb{F}_{2}$-valued trace of $\mathcal{R}$ (for instance, see \cite{Bhagat2024Trace}, Example $3.13$).
\begin{theorem}
    Let $\mathcal{B}=\{\mathbf{b}_1 = 1 + u + v, \mathbf{b}_2 = u + v, \mathbf{b}_3 = u,  \mathbf{b}_4 = uv\}$ be an $\mathbb{F}_{2}$-basis  of $\mathcal{R}$. Suppose $\mathcal{C}$ is a linear code over $\mathcal{R}$ and $G = \mathbf{b}_{1}G_{1}+\mathbf{b}_{2}G_{2}+\mathbf{b}_{3}G_{3}+\mathbf{b}_{4}G_{4}$ is its generator matrix, where $G_{i} \in M_{k\times n}(\mathbb{F}_{2})$ for $i=1,2,3,4.$ Then $\mathcal{C}^{(2)}$ is binary linear code of length $n$ and is generated by $$G^{(2)} = 
    \left[
    \begin{tabular}{c}
         $G_{1}+G_{4}$ \\
         $G_{3}$ \\
         $G_{2}$ \\
         $G_{1}$
    \end{tabular}
    \right]$$
Furthermore, if $D=\mathbf{b}_{1}D_{1}+\mathbf{b}_{2}D_{2}+\mathbf{b}_{3}D_{3}+\mathbf{b}_{4}D_{4}$, then $\mathcal{C}_{D}^{(2)} = \mathcal{C}_{D^{(2)}}$, where $D^{(2)} = \{ (d_{1}+d_{4},d_{3},d_{2},d_{1}) : d_{i} \in D_{i}, 1\leq i\leq 4\}.$
\end{theorem}
\begin{proof}
    Let $\mathbf{g}_{jk} = \mathbf{g}_{jk}^{(1)}\mathbf{b}_{1} + \mathbf{g}_{jk}^{(2)}\mathbf{b}_{2}+\mathbf{g}_{jk}^{(3)}\mathbf{b}_{3} + \mathbf{g}_{jk}^{(4)}\mathbf{b}_{4}$, where $\mathbf{g}_{jk}^{(l)} \in \mathbb{F}_{2}$ for $l=1,2,3,4$. Then 
\begin{eqnarray*}
\tau(\mathbf{g}_{jk}\mathbf{b}_{1}) &=&  \mathbf{g}_{jk}^{(1)} + \mathbf{g}_{jk}^{(4)}, \\
\tau(\mathbf{g}_{jk}\mathbf{b}_{2}) &=& \mathbf{g}_{jk}^{(3)}, \\
\tau(\mathbf{g}_{jk}\mathbf{b}_{3}) &=& \mathbf{g}_{jk}^{(2)}, \\
\tau(\mathbf{g}_{jk}\mathbf{b}_{4}) &=& \mathbf{g}_{jk}^{(1)}.
\end{eqnarray*} 
Using Theorem \ref{trace_matrix}, we have the desired result.  \qed
\end{proof}

Consider the map $c_{D}^{(2)}: (\mathbb{F}_{2}^{m})^{4} \to \mathcal{C}_{D}^{(2)}$ defined by $\mathbf{v} \mapsto (\mathbf{v}\cdot \mathbf{d})_{\mathbf{d}\in D^{(2)}}$. Clearly, the map $c_{D}^{(2)}$ is a surjective linear transformation. Thus 
$$\mathcal{C}_{D}^{(2)} = \{ c_{D}^{(2)}(\mathbf{x}_{1}, \mathbf{x}_{2}, \mathbf{x}_{3},\mathbf{x}_{4}) = (\mathbf{x}_{1}, \mathbf{x}_{2}, \mathbf{x}_{3},\mathbf{x}_{4})\cdot (d_{1}+d_{4},d_{3},d_{2},d_{1})_{d_i \in D_{i}} : \mathbf{x}_{1}, \mathbf{x}_{2}, \mathbf{x}_{3}, \mathbf{x}_{4} \in \mathbb{F}_{2}^{m}\}.$$
Now,
\begin{eqnarray} \label{weight_equation}
    \wt\left(c_{D}^{(2)}(\mathbf{x}_{1}, \mathbf{x}_{2}, \mathbf{x}_{3},\mathbf{x}_{4})\right) &=& \wt\left( (\mathbf{x}_{1}, \mathbf{x}_{2}, \mathbf{x}_{3},\mathbf{x}_{4})\cdot (d_{1}+d_{4},d_{3},d_{2},d_{1})_{d_i \in D_{i}}\right) \nonumber\\
    &=& \wt\left( ((\mathbf{x}_{1}+\mathbf{x}_{4})d_{1}+\mathbf{x}_{3} d_{2}+ \mathbf{x}_{2} d_{3} + \mathbf{x}_{1} d_{4} )_{d_{i}\in D_{i}} \right) \nonumber\\ 
    &=& |D| -\frac{1}{2}\sum\limits_{d_{1}\in D_{1}}\sum\limits_{d_{2}\in D_{2}}\sum\limits_{d_{3}\in D_{3}}\sum\limits_{d_{4}\in D_{4}} \left( 1+(-1)^{(\mathbf{x}_{1}+\mathbf{x}_{4})d_{1}+\mathbf{x}_{3} d_{2}+\mathbf{x}_{2} d_{3}+\mathbf{x}_{1} d_{4}} \right)  \\
    &=& \frac{|D|}{2} - \frac{1}{2} \sum\limits_{d_{1}\in D_{1}} (-1)^{(\mathbf{x}_{1}+\mathbf{x}_{4}) d_{1}} \sum\limits_{d_{2}\in D_{2}} (-1)^{\mathbf{x}_{3} d_{2}} \sum\limits_{d_{3}\in D_{3}} (-1)^{\mathbf{x}_{2} d_{3}} \sum\limits_{d_{4}\in D_{4}} (-1)^{\mathbf{x}_{1} d_{4}}. \nonumber
\end{eqnarray}
Let $\mathbf{x}_{1} \in \mathbb{F}_{2}^{m}$ and $\emptyset \neq Y \subseteq [m]$. Define a Boolean function $\psi(\cdot\mid Y) : \mathbb{F}_{2}^{m} \to \mathbb{F}_{2}$ by \[
\psi(\mathbf{x}_{1} \mid Y) := \prod_{i \in Y} (1 - \alpha_i)
= 
\begin{cases}
1, & \text{if } \Supp(\mathbf{x}_{1}) \cap Y = \emptyset, \\
0, & \text{if } \Supp(\mathbf{x}_{1}) \cap Y \neq \emptyset.
\end{cases}
\]
Then,
\begin{eqnarray}\label{boolean_relation}
 \sum\limits_{t \in \Delta_{Y}} (-1)^{\mathbf{x}_{1} t} &=& \mathcal{H}_{\Delta_{Y}} ((-1)^{\alpha_{1}},(-1)^{\alpha_{2}},\ldots,(-1)^{\alpha_{m}}) \nonumber \\
 &=& \prod\limits_{i\in Y} \left( 1+(-1)^{\alpha_{i}}\right) = \prod\limits_{i \in Y} (2-2\alpha_{i}) \nonumber \\
 &=& 2^{|Y|}\prod\limits_{i\in Y}(1-\alpha_{i}) = 2^{|Y|}\psi(\mathbf{x}_{1} \mid Y).
\end{eqnarray}
\begin{lemma}\cite{shi2022few-weight}\label{lemma_complement}
    For $\mathbf{x}_{1} \in \mathbb{F}_{2}^{m},$ we have 
    \[ \sum\limits_{t \in \Delta_{Y}^{c}} (-1)^{\mathbf{x}_{1} t} = 2^{m}\delta_{0,\mathbf{x}_{1}} - 2^{|Y|}\psi(\mathbf{x}_{1} \mid Y), \]
    where $\Delta_{Y}^{c} = \mathbb{F}_{2}^{m}\setminus \Delta_{Y}$.
\end{lemma}
\section{Hamming Weight Distribution of $\mathcal{C}_D^{(2)}$}\label{Section 4}
Here, we find the Hamming weight distributions of $\mathcal{C}_D^{(2)}$. The following lemma is helpful to prove the main result of this section.
\begin{lemma}
    Let $X$ and $W$ be two subsets of $[m]$. Then
    \begin{enumerate}
        \item
        \begin{enumerate}
            \item $|\{\mathbf{x}\in \mathbb{F}_{2}^{m}\setminus\{0\}:\psi(\mathbf{x}\mid X) = 0\}| = (2^{|X|}-1)2^{m-|X|} = 2^{m}-2^{m-|X|}.$
            \item $|\{\mathbf{x}\in \mathbb{F}_{2}^{m}\setminus\{0\}: \psi(\mathbf{x}\mid X) = 1\}| = 2^{m-|X|}-1.$
        \end{enumerate}
        \item 
        \begin{enumerate}
            \item $|\{\mathbf{x}\in \mathbb{F}_{2}^{m}\setminus\{0\}:\psi(\mathbf{x}\mid X)=0, \psi(\mathbf{x}\mid W) = 0\}| = (2^{|X|}-1)2^{m-|X|}+(2^{|W|}-1)2^{m-|W|}-(2^{|X\cup W|}-1)2^{m-|X\cup W|} = 2^{m}-2^{m-|X|}-2^{m-|W|}+2^{m-|X\cup W|} .$
            \item $|\{\mathbf{x}\in \mathbb{F}_{2}^{m}\setminus\{0\}:\psi(\mathbf{x}\mid X)=1, \psi(\mathbf{x}\mid W) = 0\}| = (2^{|W\setminus X|}-1)2^{m-|X\cup W|} = 2^{m-|X|}-2^{m-|X\cup W|}.$
            \item $|\{\mathbf{x}\in \mathbb{F}_{2}^{m}\setminus\{0\}:\psi(\mathbf{x}\mid X)=0, \psi(\mathbf{x}\mid W) = 1\}| = (2^{|X\setminus W|}-1)2^{m-|X\cup W|} = 2^{m-|W|}-2^{m-|X\cup W|}.$
            \item $|\{\mathbf{x}\in \mathbb{F}_{2}^{m}\setminus\{0\}:\psi(\mathbf{x}\mid X)=1, \psi(\mathbf{x}\mid W) = 1\}| = 2^{m-|X\cup W|}-1.$
        \end{enumerate}
        \item 
        \begin{enumerate}
            \item $|\{(\mathbf{x},\mathbf{y})\in \mathbb{F}_{2}^{m}\times \mathbb{F}_{2}^{m}\setminus\{(0,0)\}:\mathbf{x} \neq \mathbf{y} ,\mathbf{y} \cap X = \emptyset, \psi(\mathbf{x}+\mathbf{y}\mid X)=0, \psi(\mathbf{x}\mid W) = 0\}| = \left((2^{|X|}-1)2^{m-|X|}+(2^{|W|}-1)2^{m-|W|}-(2^{|X\cup W|}-1)2^{m-|X\cup W|}\right) \times \left(2^{m-|X|}-1 \right).$
            \item $|\{(\mathbf{x},\mathbf{y})\in \mathbb{F}_{2}^{m}\times \mathbb{F}_{2}^{m}\setminus\{(0,0)\}:\mathbf{x} \neq \mathbf{y} ,\mathbf{x}\cap X = \emptyset, \mathbf{y} \cap X \neq \emptyset, \psi(\mathbf{x}+\mathbf{y}\mid X)=0, \psi(\mathbf{x}\mid W) = 0\}| = \left((2^{|W\setminus X|}-1)2^{m-|X\cup W|}\right) \times \left((2^{|X|}-1)2^{m-|X|} \right).$
            \item $|\{(\mathbf{x},\mathbf{y})\in \mathbb{F}_{2}^{m}\times \mathbb{F}_{2}^{m}\setminus\{(0,0)\}:\mathbf{x} \neq \mathbf{y} ,\mathbf{x}\cap X \neq \emptyset, \mathbf{y} \cap X \neq \emptyset, \psi(\mathbf{x}+\mathbf{y}\mid X)=0, \psi(\mathbf{x}\mid W) = 0\}| = \left((2^{|X|}-1)2^{m-|X|}+(2^{|W|}-1)2^{m-|W|}-(2^{|X\cup W|}-1)2^{m-|X\cup W|}\right) \times \\ \left((2^{|X|}-2)2^{m-|X|} \right).$
        \end{enumerate}
        \item
        \begin{enumerate}
            \item $|\{(\mathbf{x},\mathbf{y})\in \mathbb{F}_{2}^{m}\times \mathbb{F}_{2}^{m}\setminus\{(0,0)\}:\mathbf{x} \neq \mathbf{y} , \mathbf{y} \cap X = \emptyset, \psi(\mathbf{x}+\mathbf{y}\mid X)=1, \psi(\mathbf{x}\mid W) = 0\}| = \left((2^{|W\setminus X|}-1)2^{m-|X\cup W|}\right) \times \left(2^{m-|X|}-2 \right).$
            \item $|\{(\mathbf{x},\mathbf{y})\in \mathbb{F}_{2}^{m}\times \mathbb{F}_{2}^{m}\setminus\{(0,0)\}:\mathbf{x} \neq \mathbf{y}, \mathbf{y} \cap X \neq \emptyset, \psi(\mathbf{x}+\mathbf{y}\mid X)=1, \psi(\mathbf{x}\mid W) = 0\}| = \left((2^{|X|}-1)2^{m-|X|}+(2^{|W|}-1)2^{m-|W|}-(2^{|X\cup W|}-1)2^{m-|X\cup W|}\right) \times \left(2^{m-|X|}-1 \right).$
        \end{enumerate}
        \item \begin{enumerate}
            \item $|\{(\mathbf{x},\mathbf{y})\in \mathbb{F}_{2}^{m}\times \mathbb{F}_{2}^{m}\setminus\{(0,0)\}:\mathbf{x} \neq \mathbf{y}, \mathbf{y} \cap X = \emptyset, \psi(\mathbf{x}+\mathbf{y}\mid X)=0, \psi(\mathbf{x}\mid W) = 1\}| = \left((2^{|X\setminus W|}-1)2^{m-|X\cup W|}\right) \times \left(2^{m-|X|}-1 \right).$
            \item $|\{(\mathbf{x},\mathbf{y})\in \mathbb{F}_{2}^{m}\times \mathbb{F}_{2}^{m}\setminus\{(0,0)\}:\mathbf{x} \neq \mathbf{y}, \mathbf{x} \cap X = \emptyset, \mathbf{y} \cap X \neq \emptyset, \psi(\mathbf{x}+\mathbf{y}\mid X)=0, \psi(\mathbf{x}\mid W) = 1\}| = \left(2^{m-|X\cup W|}-1\right) \times \left((2^{|X|}-1)2^{m-|X|} \right).$
            \item $|\{(\mathbf{x},\mathbf{y})\in \mathbb{F}_{2}^{m}\times \mathbb{F}_{2}^{m}\setminus\{(0,0)\}:\mathbf{x} \neq \mathbf{y}, \mathbf{x} \cap X \neq \emptyset, \mathbf{y} \cap X \neq \emptyset, \psi(\mathbf{x}+\mathbf{y}\mid X)=0, \psi(\mathbf{x}\mid W) = 1\}| = \left((2^{|X\setminus W|}-1)2^{m-|X\cup W|}\right) \times \left((2^{|X|}-2)2^{m-|X|} \right).$
        \end{enumerate}
        \item 
        \begin{enumerate}
            \item $|\{(\mathbf{x},\mathbf{y})\in \mathbb{F}_{2}^{m}\times \mathbb{F}_{2}^{m}\setminus\{(0,0)\}:\mathbf{x} \neq \mathbf{y}, \mathbf{y} \cap X = \emptyset, \psi(\mathbf{x}+\mathbf{y}\mid X)=1, \psi(\mathbf{x}\mid W) = 1\}| = \left(2^{m-|X\cup W|}-1\right) \times \left(2^{m-|X|}-2 \right).$
            \item $|\{(\mathbf{x},\mathbf{y})\in \mathbb{F}_{2}^{m}\times \mathbb{F}_{2}^{m}\setminus\{(0,0)\}:\mathbf{x} \neq \mathbf{y}, \mathbf{y} \cap X \neq \emptyset, \psi(\mathbf{x}+\mathbf{y}\mid X)=1, \psi(\mathbf{x}\mid W) = 1\}| = \left((2^{|X\setminus W|}-1)2^{m-|X\cup W|}\right) \times \left(2^{m-|X|}-1 \right).$
        \end{enumerate}
    \end{enumerate}
\end{lemma}
In the following theorem, we determine the weight distributions of $\mathcal{C}_{D}^{(2)}$ for different choices of $D$.
\begin{theorem}\label{maintheorem}
    Let $m$ be a natural number.
    \begin{enumerate}
        \item For the defining set $D=\mathbf{b}_{1}\Delta_{X}+\mathbf{b}_{2}\Delta_{Y}+\mathbf{b}_{3}\Delta_{Z}+\mathbf{b}_{4}\Delta_{W} \subseteq \mathcal{R}^{m},$ $\mathcal{C}_{D}^{(2)} $ is a one-weight binary linear code with length $|D|=2^{|X|+|Y|+|Z|+|W|}$, dimension $|X|+|Y|+|Z|+|W|$ and minimum distance $2^{|X|+|Y|+|Z|+|W|-1}.$ Its Hamming weight distribution is displayed in Table \ref{table1}. Moreover, if $|X|+|Y|+|Z|+|W| \geq 2$, then $\mathcal{C}_{D}^{(2)}$ is distance-optimal. 
        \item For the defining set $D=\mathbf{b}_{1}\Delta_{X}^{c}+\mathbf{b}_{2}\Delta_{Y}+\mathbf{b}_{3}\Delta_{Z}+\mathbf{b}_{4}\Delta_{W} \subseteq \mathcal{R}^{m}$ with $|X|< m,$ $\mathcal{C}_{D}^{(2)} $ is a two-weight binary linear code with length $|D|=(2^m-2^{|X|})2^{|Y|+|Z|+|W|}$, dimension $m+|Y|+|Z|+|W|$ and minimum distance $(2^m-2^{|X|})2^{|Y|+|Z|+|W|-1}.$ Its Hamming weight distribution is displayed in Table \ref{table2}. In addition, $\mathcal{C}_{D}^{(2)}$ attains the Griesmer bound and is therefore distance-optimal. 
         \item For the defining set $D=\mathbf{b}_{1}\Delta_{X}^{c}+\mathbf{b}_{2}\Delta_{Y}^{c}+\mathbf{b}_{3}\Delta_{Z}+\mathbf{b}_{4}\Delta_{W}\subseteq \mathcal{R}^{m}$ with $|X| \neq |Y|$ and $|X|,|Y| < m$, $\mathcal{C}_{D}^{(2)} $ is a four-weight binary linear code with length $|D|=(2^m-2^{|X|})(2^m-2^{|Y|})2^{|Z|+|W|}$, dimension $2m+|Z|+|W|$ and minimum distance $(2^m-2^{|X|}-2^{|Y|})2^{m+|Z|+|W|-1}.$ Its Hamming weight distribution is displayed in Table \ref{table6}. Moreover, if $2^{|X|+|Y|+|Z|+|W|} \leq m+|Z|+|W|+\min\{|X|,|Y|\}$, then $\mathcal{C}_{D}^{(2)}$ is distance-optimal.
          \item For the defining set $D=\mathbf{b}_{1}\Delta_{X}^{c}+\mathbf{b}_{2}\Delta_{Y}^{c}+\mathbf{b}_{3}\Delta_{Z}^{c}+\mathbf{b}_{4}\Delta_{W}\subseteq \mathcal{R}^{m}$ with $|X|,|Y|$ and $|Z|$ all distinct and $|X|,|Y|,|Z| < m$, $\mathcal{C}_{D}^{(2)} $ is a eight-weight binary linear code with length $|D|=(2^m-2^{|X|})(2^m-2^{|Y|})(2^m-2^{|Z|})2^{|W|}$, dimension $3m+|W|$ and minimum distance $(2^m-2^{s_{1}})(2^{m}-2^{s_2}-2^{s_3})2^{m+|W|-1}$, where $s_{1} < s_{2} < s_{3}$ and $\{s_1, s_2, s_3\} = \{|X|,|Y|,|Z|\}$. The weight distribution is displayed in the Table \ref{table12}.
         \item For the defining set $D=\mathbf{b}_{1}\Delta_{X}^{c}+\mathbf{b}_{2}\Delta_{Y}^{c}+\mathbf{b}_{3}\Delta_{Z}^{c}+\mathbf{b}_{4}\Delta_{W}^{c}\subseteq \mathcal{R}^{m}$ with $|X|,|Y|,|Z|$ and $|W|$ all distinct and $|X|,|Y|,|Z|,|W| < m$, $\mathcal{C}_{D}^{(2)} $ is a sixteen-weight binary linear code with length $|D|=(2^m-2^{|X|})(2^m-2^{|Y|})(2^m-2^{|Z|})(2^m-2^{|W|})$, dimension $4m$ and minimum distance $(2^m-2^{s_{1}})(2^m-2^{s_2})(2^{m}-2^{s_3}-2^{s_4})2^{m-1}$, where $s_{1}< s_{2} < s_{3} < s_{4}$ and $\{s_1, s_2, s_3,s_4\} = \{|X|,|Y|,|Z|,|W|\}$. The Hamming weight distribution is displayed in the Table \ref{table16}.
         \item For the defining set $D=\mathbf{b}_{1}\Delta_{X}+\mathbf{b}_{2}\Delta_{Y}+\mathbf{b}_{3}\Delta_{Z}+\mathbf{b}_{4}\Delta_{W}\subseteq \mathcal{R}^{m},$ $\mathcal{C}_{D^c}^{(2)} $ is a two-weight binary linear code with length $|D^c|=2^{4m}-2^{|X|+|Y|+|Z|+|W|}$, dimension $4m$ and minimum distance $2^{4m-1}-2^{|X|+|Y|+|Z|+|W|-1}.$ Its Hamming weight distribution is displayed in Table \ref{table17}. In addition, $C_{D^c}^{(2)}$ attains the Griesmer bound and is therefore distance-optimal.
    \end{enumerate}
\end{theorem}
\begin{proof}
    Now, we will prove part $4$, and the other cases are similar to it. Using Lemma \ref{lemma_complement} and Equations \eqref{weight_equation} and \eqref{boolean_relation}, we have
   \begin{align*}
       \wt(c_{D}^{(2)}(\mathbf{x}_{1},\mathbf{x}_{2},\mathbf{x}_{3},\mathbf{x}_{4})) &= \frac{|D|}{2}-\frac{1}{2}\left\{ (2^{m}\delta_{0,\mathbf{x}_{1}+\mathbf{x}_{4}}-2^{|X|}\psi(\mathbf{x}_{1}+\mathbf{x}_{4}\mid X)) (2^{m}\delta_{0,\mathbf{x}_{3}}-2^{|Y|}\psi(\mathbf{x}_{3}\mid Y)) \right. \\&\left.(2^{m}\delta_{0,\mathbf{x}_{2}}-2^{|Z|}\psi(\mathbf{x}_{2}\mid Z)) 2^{|W|}\psi(\mathbf{x}_{1}\mid W) \right\}
   \end{align*} 
The $16$ possible cases are as follows:
\begin{enumerate}
    \item[(1)] $\boldsymbol{\mathbf{x}_{1} + \mathbf{x}_{4} = 0, \; \mathbf{x}_{3} = 0, \; \mathbf{x}_{2} = 0, \; \mathbf{x}_{1} = 0}$
  $ \implies \mathbf{x}_{1}=0, \mathbf{x}_{2} = 0, \mathbf{x}_{3} = 0, \mathbf{x}_{4} = 0$. Then
\begin{equation*}
    \wt(c_{D}^{(2)}(\mathbf{x}_{1},\mathbf{x}_{2},\mathbf{x}_{3},\mathbf{x}_{4})) = 0
\end{equation*}
In this case, $\#(\mathbf{x}_{1},\mathbf{x}_{2},\mathbf{x}_{3},\mathbf{x}_{4}) = 1$.
\item[(2)] $\boldsymbol{\mathbf{x}_{1}+\mathbf{x}_{4} \neq 0, \mathbf{x}_{3} = 0, \mathbf{x}_{2} = 0, \mathbf{x}_{1} = 0} \implies \mathbf{x}_{1} = 0, \mathbf{x}_{2} = 0, \mathbf{x}_{3} = 0, \mathbf{x}_{4} \neq 0.$ Then
\begin{equation*}
    \wt(c_{D}^{(2)}(\mathbf{x}_{1},\mathbf{x}_{2},\mathbf{x}_{3},\mathbf{x}_{4})) = \frac{|D|}{2}+\frac{1}{2}(2^{m}-2^{|Y|})(2^{m}-2^{|Z|})2^{|X|+|W|}\psi(\mathbf{x}_{4}\mid X)
\end{equation*}
    \begin{enumerate}
        \item If $\psi(\mathbf{x}_{4}\mid X) = 0$, then $\wt(c_{D}^{(2)}(\mathbf{x}_{1},\mathbf{x}_{2},\mathbf{x}_{3},\mathbf{x}_{4})) = \frac{|D|}{2}.$ 
        
        In this case, $\#(\mathbf{x}_{1},\mathbf{x}_{2},\mathbf{x}_{3},\mathbf{x}_{4}) = 2^{m}-2^{m-|X|}.$
        \item If $\psi(\mathbf{x}_{4}\mid X) = 1$, then $\wt(c_{D}^{(2)}(\mathbf{x}_{1},\mathbf{x}_{2},\mathbf{x}_{3},\mathbf{x}_{4})) = (2^{m}-2^{|Y|})(2^{m}-2^{|Z|})2^{m+|W|-1}$.

        In this case, $\#(\mathbf{x}_{1},\mathbf{x}_{2},\mathbf{x}_{3},\mathbf{x}_{4}) = 2^{m-|X|}-1$
    \end{enumerate}
    
\item[(3)] $\boldsymbol{\mathbf{x}_{1}+\mathbf{x}_{4} = 0, \mathbf{x}_{3} \neq 0, \mathbf{x}_{2} = 0, \mathbf{x}_{1} = 0} \implies \mathbf{x}_{1} = 0, \mathbf{x}_{2} = 0, \mathbf{x}_{3} \neq 0, \mathbf{x}_{4} = 0.$ Then
\begin{equation*}
    \wt(c_{D}^{(2)}(\mathbf{x}_{1},\mathbf{x}_{2},\mathbf{x}_{3},\mathbf{x}_{4})) = \frac{|D|}{2}+\frac{1}{2}(2^{m}-2^{|X|})(2^{m}-2^{|Z|})2^{|Y|+|W|}\psi(\mathbf{x}_{3}\mid Y)
\end{equation*}
    \begin{enumerate}
        \item If $\psi(\mathbf{x}_{3}\mid Y) = 0$, then $\wt(c_{D}^{(2)}(\mathbf{x}_{1},\mathbf{x}_{2},\mathbf{x}_{3},\mathbf{x}_{4})) = \frac{|D|}{2}.$ 
        
        In this case, $\#(\mathbf{x}_{1},\mathbf{x}_{2},\mathbf{x}_{3},\mathbf{x}_{4}) = 2^{m}-2^{m-|Y|}$
        \item If $\psi(\mathbf{x}_{3}\mid Y) = 1$, then $\wt(c_{D}^{(2)}(\mathbf{x}_{1},\mathbf{x}_{2},\mathbf{x}_{3},\mathbf{x}_{4})) = (2^{m}-2^{|X|})(2^{m}-2^{|Z|})2^{m+|W|-1}$.

        In this case, $\#(\mathbf{x}_{1},\mathbf{x}_{2},\mathbf{x}_{3},\mathbf{x}_{4}) = 2^{m-|Y|}-1$
    \end{enumerate}
    \item[(4)] $\boldsymbol{\mathbf{x}_{1}+\mathbf{x}_{4} = 0, \mathbf{x}_{3} = 0, \mathbf{x}_{2} \neq 0, \mathbf{x}_{1} = 0} \implies \mathbf{x}_{1} = 0, \mathbf{x}_{2} \neq 0, \mathbf{x}_{3} = 0, \mathbf{x}_{4} = 0.$ Then
\begin{equation*}
    \wt(c_{D}^{(2)}(\mathbf{x}_{1},\mathbf{x}_{2},\mathbf{x}_{3},\mathbf{x}_{4})) = \frac{|D|}{2}+\frac{1}{2}(2^{m}-2^{|X|})(2^{m}-2^{|Y|})2^{|Z|+|W|}\psi(\mathbf{x}_{2}\mid Z)
\end{equation*}
    \begin{enumerate}
        \item If $\psi(\mathbf{x}_{2}\mid Z) = 0$, then $\wt(c_{D}^{(2)}(\mathbf{x}_{1},\mathbf{x}_{2},\mathbf{x}_{3},\mathbf{x}_{4})) = \frac{|D|}{2}.$ 
        
        In this case, $\#(\mathbf{x}_{1},\mathbf{x}_{2},\mathbf{x}_{3},\mathbf{x}_{4}) = 2^{m}-2^{m-|Z|}$
        \item If $\psi(\mathbf{x}_{2}\mid Z) = 1$, then $\wt(c_{D}^{(2)}(\mathbf{x}_{1},\mathbf{x}_{2},\mathbf{x}_{3},\mathbf{x}_{4})) = (2^{m}-2^{|X|})(2^{m}-2^{|Y|})2^{m+|W|-1}$.

        In this case, $\#(\mathbf{x}_{1},\mathbf{x}_{2},\mathbf{x}_{3},\mathbf{x}_{4}) = 2^{m-|Z|}-1$
    \end{enumerate}
    \item[(5)] $\boldsymbol{\mathbf{x}_{1}+\mathbf{x}_{4} = 0, \mathbf{x}_{3} = 0, \mathbf{x}_{2} = 0, \mathbf{x}_{1} \neq 0} \implies \mathbf{x}_{1} = \mathbf{x}_{4} \neq 0, \mathbf{x}_{2} = 0, \mathbf{x}_{3} = 0.$ Then
\begin{equation*}
    \wt(c_{D}^{(2)}(\mathbf{x}_{1},\mathbf{x}_{2},\mathbf{x}_{3},\mathbf{x}_{4})) = \frac{|D|}{2}-\frac{|D|}{2}\psi(\mathbf{x}_{1}\mid W)
\end{equation*}
    \begin{enumerate}
        \item If $\psi(\mathbf{x}_{1}\mid W) = 0$, then $\wt(c_{D}^{(2)}(\mathbf{x}_{1},\mathbf{x}_{2},\mathbf{x}_{3},\mathbf{x}_{4})) = \frac{|D|}{2}.$ 
        
        In this case, $\#(\mathbf{x}_{1},\mathbf{x}_{2},\mathbf{x}_{3},\mathbf{x}_{4}) = 2^{m}-2^{m-|W|}$
        \item If $\psi(\mathbf{x}_{1}\mid W) = 1$, then $\wt(c_{D}^{(2)}(\mathbf{x}_{1},\mathbf{x}_{2},\mathbf{x}_{3},\mathbf{x}_{4})) = 0$.

        In this case, $\#(\mathbf{x}_{1},\mathbf{x}_{2},\mathbf{x}_{3},\mathbf{x}_{4}) = 2^{m-|W|}-1$
    \end{enumerate} 
    \item[(6)] 
$\boldsymbol{\mathbf{x}_{1}+\mathbf{x}_{4} \neq 0, \mathbf{x}_{3} \neq 0, \mathbf{x}_{2} = 0, \mathbf{x}_{1} = 0} \implies \mathbf{x}_{1} = 0, \mathbf{x}_{2} = 0, \mathbf{x}_{3} \neq 0, \mathbf{x}_{4} \neq 0.$ Then
\begin{equation*}
    \wt(c_{D}^{(2)}(\mathbf{x}_{1},\mathbf{x}_{2},\mathbf{x}_{3},\mathbf{x}_{4})) = \frac{|D|}{2}-\frac{1}{2}(2^{m}-2^{|Z|})2^{|X|+|Y|+|W|}\psi(\mathbf{x}_{4}\mid X)\psi(\mathbf{x}_{3}\mid Y)
\end{equation*}
    \begin{enumerate}
        \item If $\psi(\mathbf{x}_{4}\mid X) = 0$ and $\psi(\mathbf{x}_{3}\mid Y) = 0$, then $\wt(c_{D}^{(2)}(\mathbf{x}_{1},\mathbf{x}_{2},\mathbf{x}_{3},\mathbf{x}_{4})) = \frac{|D|}{2}.$ 
        
        In this case, $\#(\mathbf{x}_{1},\mathbf{x}_{2},\mathbf{x}_{3},\mathbf{x}_{4}) = \left( (2^{|Y|}-1)2^{m-|Y|}\right) \times  \left((2^{|X|}-1)2^{m-|X|}\right)$.
        \item If $\psi(\mathbf{x}_{4}\mid X) = 1$ and $\psi(\mathbf{x}_{3}\mid Y) = 0$, then $\wt(c_{D}^{(2)}(\mathbf{x}_{1},\mathbf{x}_{2},\mathbf{x}_{3},\mathbf{x}_{4})) = \frac{|D|}{2}.$ 
        
        In this case, $\#(\mathbf{x}_{1},\mathbf{x}_{2},\mathbf{x}_{3},\mathbf{x}_{4}) = \left((2^{|Y|}-1)2^{m-|Y|}\right) \times \left(2^{m-|X|}-1\right)$.
        \item If $\psi(\mathbf{x}_{4}\mid X) = 0$ and $\psi(\mathbf{x}_{3}\mid Y) = 1$, then $\wt(c_{D}^{(2)}(\mathbf{x}_{1},\mathbf{x}_{2},\mathbf{x}_{3},\mathbf{x}_{4})) = \frac{|D|}{2}.$ 
        
        In this case, $\#(\mathbf{x}_{1},\mathbf{x}_{2},\mathbf{x}_{3},\mathbf{x}_{4}) = \left(2^{m-|Y|}-1 \right) \times \left((2^{|X|}-1)2^{m-|X|} \right)$.
        \item If $\psi(\mathbf{x}_{4}\mid X) = 1$ and $\psi(\mathbf{x}_{3}\mid Y) = 1$, then $\wt(c_{D}^{(2)}(\mathbf{x}_{1},\mathbf{x}_{2},\mathbf{x}_{3},\mathbf{x}_{4})) = (2^{m}-2^{|Z|})(2^{m}-2^{|X|}-2^{|Y|})2^{m+|W|-1}.$ 
        
        In this case, $\#(\mathbf{x}_{1},\mathbf{x}_{2},\mathbf{x}_{3},\mathbf{x}_{4}) = \left( 2^{m-|Y|}-1\right) \times \left(2^{m-|X|}-1 \right)$.
    \end{enumerate} 

    \item[(7)] 
$\boldsymbol{\mathbf{x}_{1}+\mathbf{x}_{4} \neq 0, \mathbf{x}_{3} = 0, \mathbf{x}_{2} \neq 0, \mathbf{x}_{1} = 0} \implies \mathbf{x}_{1} = 0, \mathbf{x}_{2} \neq 0, \mathbf{x}_{3} = 0, \mathbf{x}_{4} \neq 0.$ Then
\begin{equation*}
    \wt(c_{D}^{(2)}(\mathbf{x}_{1},\mathbf{x}_{2},\mathbf{x}_{3},\mathbf{x}_{4})) = \frac{|D|}{2}-\frac{1}{2}(2^{m}-2^{|Y|})2^{|X|+|Z|+|W|}\psi(\mathbf{x}_{4}\mid X)\psi(\mathbf{x}_{2}\mid Z)
\end{equation*}
    \begin{enumerate}
        \item If $\psi(\mathbf{x}_{4}\mid X) = 0$ and $\psi(\mathbf{x}_{2}\mid Z) = 0$, then $\wt(c_{D}^{(2)}(\mathbf{x}_{1},\mathbf{x}_{2},\mathbf{x}_{3},\mathbf{x}_{4})) = \frac{|D|}{2}.$ 
        
        In this case, $\#(\mathbf{x}_{1},\mathbf{x}_{2},\mathbf{x}_{3},\mathbf{x}_{4}) = \left((2^{|Z|}-1)2^{m-|Z|} \right) \times \left((2^{|X|}-1)2^{m-|X|} \right)$.
        \item If $\psi(\mathbf{x}_{4}\mid X) = 1$ and $\psi(\mathbf{x}_{2}\mid Z) = 0$, then $\wt(c_{D}^{(2)}(\mathbf{x}_{1},\mathbf{x}_{2},\mathbf{x}_{3},\mathbf{x}_{4})) = \frac{|D|}{2}.$ 
        
        In this case, $\#(\mathbf{x}_{1},\mathbf{x}_{2},\mathbf{x}_{3},\mathbf{x}_{4}) = \left( (2^{|Z|}-1)2^{m-|Z|}\right) \times \left(2^{m-|X|}-1 \right)$.
        \item If $\psi(\mathbf{x}_{4}\mid X) = 0$ and $\psi(\mathbf{x}_{2}\mid Z) = 1$, then $\wt(c_{D}^{(2)}(\mathbf{x}_{1},\mathbf{x}_{2},\mathbf{x}_{3},\mathbf{x}_{4})) = \frac{|D|}{2}.$ 
        
        In this case, $\#(\mathbf{x}_{1},\mathbf{x}_{2},\mathbf{x}_{3},\mathbf{x}_{4}) = \left(2^{m-|Z|}-1 \right) \times \left((2^{|X|}-1)2^{m-|X|} \right)$.
        \item If $\psi(\mathbf{x}_{4}\mid X) = 1$ and $\psi(\mathbf{x}_{2}\mid Z) = 1$, then $\wt(c_{D}^{(2)}(\mathbf{x}_{1},\mathbf{x}_{2},\mathbf{x}_{3},\mathbf{x}_{4})) = (2^{m}-2^{|Y|})(2^{m}-2^{|X|}-2^{|Z|})2^{m+|W|-1}.$ 
        
        In this case, $\#(\mathbf{x}_{1},\mathbf{x}_{2},\mathbf{x}_{3},\mathbf{x}_{4}) = \left(2^{m-|Z|}-1 \right) \times \left(2^{m-|X|}-1 \right)$.
    \end{enumerate} 
    \item[(8)] 
$\boldsymbol{\mathbf{x}_{1}+\mathbf{x}_{4} \neq 0, \mathbf{x}_{3} = 0, \mathbf{x}_{2} = 0, \mathbf{x}_{1} \neq 0}.$ Then
\begin{equation*}
    \wt(c_{D}^{(2)}(\mathbf{x}_{1},\mathbf{x}_{2},\mathbf{x}_{3},\mathbf{x}_{4})) = \frac{|D|}{2}+\frac{1}{2}(2^{m}-2^{|Y|})(2^{m}-2^{|Z|})2^{|X|+|W|}\psi(\mathbf{x}_{1}+\mathbf{x}_{4}\mid X)\psi(\mathbf{x}_{1}\mid W)
\end{equation*}
    Subcase - (I) $\mathbf{x}_{4} = 0$ 
    
    $\implies \mathbf{x}_{1} \neq 0, \mathbf{x}_{3} = 0, \mathbf{x}_{2} = 0, \mathbf{x}_{4} = 0$. Then 
\begin{equation*}
    \wt(c_{D}^{(2)}(\mathbf{x}_{1},\mathbf{x}_{2},\mathbf{x}_{3},\mathbf{x}_{4})) = \frac{|D|}{2}+\frac{1}{2}(2^{m}-2^{|Y|})(2^{m}-2^{|Z|})2^{|X|+|W|}\psi(\mathbf{x}_{1}\mid X)\psi(\mathbf{x}_{1}\mid W)
\end{equation*}
    \begin{enumerate}
        \item If $\psi(\mathbf{x}_{1}\mid X) = 0$ and $\psi(\mathbf{x}_{1}\mid W) = 0$, then $\wt(c_{D}^{(2)}(\mathbf{x}_{1},\mathbf{x}_{2},\mathbf{x}_{3},\mathbf{x}_{4})) = \frac{|D|}{2}.$ 
        
        In this case, $\#(\mathbf{x}_{1},\mathbf{x}_{2},\mathbf{x}_{3},\mathbf{x}_{4}) = (2^{|X|}-1)2^{m-|X|}+(2^{|W|}-1)2^{m-|W|}-(2^{|X\cup W|}-1)2^{m-|X\cup W|}$.
        \item If $\psi(\mathbf{x}_{1}\mid X) = 1$ and $\psi(\mathbf{x}_{1}\mid W) = 0$, then $\wt(c_{D}^{(2)}(\mathbf{x}_{1},\mathbf{x}_{2},\mathbf{x}_{3},\mathbf{x}_{4})) = \frac{|D|}{2}.$ 
        
        In this case, $\#(\mathbf{x}_{1},\mathbf{x}_{2},\mathbf{x}_{3},\mathbf{x}_{4}) = (2^{|W\setminus X|}-1)2^{m-|X\cup W|}$.
        \item If $\psi(\mathbf{x}_{1}\mid X) = 0$ and $\psi(\mathbf{x}_{1}\mid W) = 1$, then $\wt(c_{D}^{(2)}(\mathbf{x}_{1},\mathbf{x}_{2},\mathbf{x}_{3},\mathbf{x}_{4})) = \frac{|D|}{2}.$ 
        
        In this case, $\#(\mathbf{x}_{1},\mathbf{x}_{2},\mathbf{x}_{3},\mathbf{x}_{4}) = (2^{|X\setminus W|}-1)2^{m-|X\cup W|}$.
        \item If $\psi(\mathbf{x}_{1}\mid X) = 1$ and $\psi(\mathbf{x}_{1}\mid W) = 1$, then $\wt(c_{D}^{(2)}(\mathbf{x}_{1},\mathbf{x}_{2},\mathbf{x}_{3},\mathbf{x}_{4})) = (2^{m}-2^{|Y|})(2^{m}-2^{|Z|})2^{m+|W|-1}.$ 
        
        In this case, $\#(\mathbf{x}_{1},\mathbf{x}_{2},\mathbf{x}_{3},\mathbf{x}_{4}) = 2^{m-|X\cup W|}-1$.
    \end{enumerate}
Subcase - (II) $\mathbf{x}_{4} \neq 0$

$\implies \mathbf{x}_{1} \neq 0, \mathbf{x}_{2} = 0, \mathbf{x}_{3} = 0, \mathbf{x}_{4} \neq 0$ and $\mathbf{x}_{1} \neq \mathbf{x}_{4}$. Then 
\begin{equation*}
    \wt(c_{D}^{(2)}(\mathbf{x}_{1},\mathbf{x}_{2},\mathbf{x}_{3},\mathbf{x}_{4})) = \frac{|D|}{2}+\frac{1}{2}(2^{m}-2^{|Y|})(2^{m}-2^{|Z|})2^{|X|+|W|}\psi(\mathbf{x}_{1}+\mathbf{x}_{4}\mid X)\psi(\mathbf{x}_{1}\mid W)
\end{equation*}
\begin{enumerate}
        \item If $\psi(\mathbf{x}_{1}+\mathbf{x}_{4}\mid X) = 0$ and $\psi(\mathbf{x}_{1}\mid W) = 0$, then $\wt(c_{D}^{(2)}(\mathbf{x}_{1},\mathbf{x}_{2},\mathbf{x}_{3},\mathbf{x}_{4})) = \frac{|D|}{2}.$
\begin{enumerate}
    \item $\mathbf{x}_{4} \cap X = \emptyset$

In this case, $ \#(\mathbf{x}_{1},\mathbf{x}_{2},\mathbf{x}_{3},\mathbf{x}_{4}) = \left(2^{m}-2^{m-|X|}-2^{m-|W|}+2^{m-|X\cup W|} \right) \times \left(2^{m-|X|}-1 \right)$.
\item $\mathbf{x}_{4} \cap X \neq \emptyset$
\begin{itemize}
    \item $\mathbf{x}_{1} \cap X = \emptyset$

In this case, $ \#(\mathbf{x}_{1},\mathbf{x}_{2},\mathbf{x}_{3},\mathbf{x}_{4}) = \left((2^{|W\setminus X|}-1)2^{m-|X\cup W|} \right) \times \left((2^{|X|}-1)2^{m-|X|} \right)$. 
\item $\mathbf{x}_{1} \cap X \neq \emptyset$

In this case, $\#(\mathbf{x}_{1},\mathbf{x}_{2},\mathbf{x}_{3},\mathbf{x}_{4}) =\left(2^{m}-2^{m-|X|}-2^{m-|W|}+2^{m-|X\cup W|} \right) \times \\ \left( (2^{|X|}-2)2^{m-|X|}\right)$.
\end{itemize}
\end{enumerate} 
        \item If $\psi(\mathbf{x}_{1}+\mathbf{x}_{4}\mid X) = 1$ and $\psi(\mathbf{x}_{1}\mid W) = 0$, then $\wt(c_{D}^{(2)}(\mathbf{x}_{1},\mathbf{x}_{2},\mathbf{x}_{3},\mathbf{x}_{4})) = \frac{|D|}{2}.$ 
        \begin{enumerate}
            \item $\mathbf{x}_{4} \cap X = \emptyset$

            In this case, $\#(\mathbf{x}_{1},\mathbf{x}_{2},\mathbf{x}_{3},\mathbf{x}_{4}) =\left( (2^{|W\setminus X|}-1)2^{m-|X\cup W|}\right) \times \left(2^{m-|X|}-2 \right).$
            \item $\mathbf{x}_{4} \cap X \neq \emptyset $

             In this case, $\#(\mathbf{x}_{1},\mathbf{x}_{2},\mathbf{x}_{3},\mathbf{x}_{4})= \left(2^{m}-2^{m-|X|}-2^{m-|W|}+2^{m-|X\cup W|}\right) \times \left(2^{m-|X|}-1 \right).$
        \end{enumerate}
        \item If $\psi(\mathbf{x}_{1}+\mathbf{x}_{4}\mid X) = 0$ and $\psi(\mathbf{x}_{1}\mid W) = 1$, then $\wt(c_{D}^{(2)}(\mathbf{x}_{1},\mathbf{x}_{2},\mathbf{x}_{3},\mathbf{x}_{4})) = \frac{|D|}{2}.$ 
        \begin{enumerate}
    \item $\mathbf{x}_{4} \cap X = \emptyset$

In this case, $\#(\mathbf{x}_{1},\mathbf{x}_{2},\mathbf{x}_{3},\mathbf{x}_{4})=\left((2^{|X\setminus W|}-1)2^{m-|X\cup W|} \right) \times \left(2^{m-|X|}-1 \right)$.
\item $\mathbf{x}_{4} \cap X \neq \emptyset$
\begin{itemize}
    \item $\mathbf{x}_{1} \cap X = \emptyset$

In this case, $\#(\mathbf{x}_{1},\mathbf{x}_{2},\mathbf{x}_{3},\mathbf{x}_{4})=\left(2^{m-|X\cup W|}-1 \right) \times \left((2^{|X|}-1)2^{m-|X|} \right)$. 
\item $\mathbf{x}_{1} \cap X \neq \emptyset$

In this case, $\#(\mathbf{x}_{1},\mathbf{x}_{2},\mathbf{x}_{3},\mathbf{x}_{4})=\left( (2^{|X\setminus W|}-1)2^{m-|X\cup W|}\right) \times \left((2^{|X|}-2)2^{m-|X|} \right)$.
\end{itemize}
\end{enumerate}
        \item If $\psi(\mathbf{x}_{1}+\mathbf{x}_{4}\mid X) = 1$ and $\psi(\mathbf{x}_{1}\mid W) = 1$, then $\wt(c_{D}^{(2)}(\mathbf{x}_{1},\mathbf{x}_{2},\mathbf{x}_{3},\mathbf{x}_{4})) = (2^{m}-2^{|Y|})(2^{m}-2^{|Z|})2^{m+|W|-1}.$ 
        
\begin{enumerate}
            \item $\mathbf{x}_{4} \cap X = \emptyset$

            In this case, $\#(\mathbf{x}_{1},\mathbf{x}_{2},\mathbf{x}_{3},\mathbf{x}_{4})=\left( 2^{m-|X\cup W|}-1\right) \times \left( 2^{m-|X|}-2\right)$.
            \item $\mathbf{x}_{4} \cap X \neq \emptyset $

             In this case, $\#(\mathbf{x}_{1},\mathbf{x}_{2},\mathbf{x}_{3},\mathbf{x}_{4})=\left((2^{|X\setminus W|}-1)2^{m-|X\cup W|} \right) \times \left(2^{m-|X|}-1 \right).$
        \end{enumerate}
    \end{enumerate}
    \item[(9)] 
$\boldsymbol{\mathbf{x}_{1}+\mathbf{x}_{4} = 0, \mathbf{x}_{3} \neq 0, \mathbf{x}_{2} \neq 0, \mathbf{x}_{1} = 0} \implies \mathbf{x}_{1} = 0, \mathbf{x}_{2} \neq 0, \mathbf{x}_{3} \neq 0, \mathbf{x}_{4} = 0.$ Then
\begin{equation*}
    \wt(c_{D}^{(2)}(\mathbf{x}_{1},\mathbf{x}_{2},\mathbf{x}_{3},\mathbf{x}_{4})) = \frac{|D|}{2}-\frac{1}{2}(2^{m}-2^{|X|})2^{|Y|+|Z|+|W|}\psi(\mathbf{x}_{3}\mid Y)\psi(\mathbf{x}_{2}\mid Z)
\end{equation*}
    \begin{enumerate}
        \item If $\psi(\mathbf{x}_{3}\mid Y) = 0$ and $\psi(\mathbf{x}_{2}\mid Z) = 0$, then $\wt(c_{D}^{(2)}(\mathbf{x}_{1},\mathbf{x}_{2},\mathbf{x}_{3},\mathbf{x}_{4})) = \frac{|D|}{2}.$ 
        
        In this case, $\#(\mathbf{x}_{1},\mathbf{x}_{2},\mathbf{x}_{3},\mathbf{x}_{4}) =\left( (2^{|Z|}-1)2^{m-|Z|}\right) \times \left((2^{|Y|}-1)2^{m-|Y|} \right)$.
        \item If $\psi(\mathbf{x}_{3}\mid Y) = 1$ and $\psi(\mathbf{x}_{2}\mid Z) = 0$, then $\wt(c_{D}^{(2)}(\mathbf{x}_{1},\mathbf{x}_{2},\mathbf{x}_{3},\mathbf{x}_{4})) = \frac{|D|}{2}.$ 
        
        In this case, $\#(\mathbf{x}_{1},\mathbf{x}_{2},\mathbf{x}_{3},\mathbf{x}_{4}) = \left((2^{|Z|}-1)2^{m-|Z|} \right) \times \left(2^{m-|Y|}-1 \right)$.
        \item If $\psi(\mathbf{x}_{3}\mid Y) = 0$ and $\psi(\mathbf{x}_{2}\mid Z) = 1$, then $\wt(c_{D}^{(2)}(\mathbf{x}_{1},\mathbf{x}_{2},\mathbf{x}_{3},\mathbf{x}_{4})) = \frac{|D|}{2}.$ 
        
        In this case, $\#(\mathbf{x}_{1},\mathbf{x}_{2},\mathbf{x}_{3},\mathbf{x}_{4}) = \left(2^{m-|Z|}-1 \right) \times \left((2^{|Y|}-1)2^{m-|Y|} \right) $.
        \item If $\psi(\mathbf{x}_{3}\mid Y) = 1$ and $\psi(\mathbf{x}_{2}\mid Z) = 1$, then $\wt(c_{D}^{(2)}(\mathbf{x}_{1},\mathbf{x}_{2},\mathbf{x}_{3},\mathbf{x}_{4})) = (2^{m}-2^{|X|})(2^{m}-2^{|Y|}-2^{|Z|})2^{m+|W|-1}.$ 
        
        In this case, $\#(\mathbf{x}_{1},\mathbf{x}_{2},\mathbf{x}_{3},\mathbf{x}_{4}) = \left(2^{m-|Z|}-1 \right) \times \left(2^{m-|Y|}-1 \right)$.
    \end{enumerate}
    \item[(10)] 
$\boldsymbol{\mathbf{x}_{1}+\mathbf{x}_{4} = 0, \mathbf{x}_{3} \neq 0, \mathbf{x}_{2} = 0, \mathbf{x}_{1} \neq 0} \implies \mathbf{x}_{1} \neq 0, \mathbf{x}_{2} = 0, \mathbf{x}_{3} \neq 0, \mathbf{x}_{4} = \mathbf{x}_{1}.$ Then
\begin{equation*}
    \wt(c_{D}^{(2)}(\mathbf{x}_{1},\mathbf{x}_{2},\mathbf{x}_{3},\mathbf{x}_{4})) = \frac{|D|}{2}+\frac{1}{2}(2^{m}-2^{|X|})(2^{m}-2^{|Z|})2^{|Y|+|W|}\psi(\mathbf{x}_{3}\mid Y)\psi(\mathbf{x}_{1}\mid W)
\end{equation*}
    \begin{enumerate}
        \item If $\psi(\mathbf{x}_{3}\mid Y) = 0$ and $\psi(\mathbf{x}_{1}\mid W) = 0$, then $\wt(c_{D}^{(2)}(\mathbf{x}_{1},\mathbf{x}_{2},\mathbf{x}_{3},\mathbf{x}_{4})) = \frac{|D|}{2}.$ 
        
        In this case, $\#(\mathbf{x}_{1},\mathbf{x}_{2},\mathbf{x}_{3},\mathbf{x}_{4}) = \left((2^{|W|}-1)2^{m-|W|} \right) \times \left( (2^{|Y|}-1)2^{m-|Y|}\right)$.
        \item If $\psi(\mathbf{x}_{3}\mid Y) = 1$ and $\psi(\mathbf{x}_{1}\mid W) = 0$, then $\wt(c_{D}^{(2)}(\mathbf{x}_{1},\mathbf{x}_{2},\mathbf{x}_{3},\mathbf{x}_{4})) = \frac{|D|}{2}.$ 
        
        In this case, $\#(\mathbf{x}_{1},\mathbf{x}_{2},\mathbf{x}_{3},\mathbf{x}_{4}) = \left((2^{|W|}-1)2^{m-|W|} \right) \times \left( 2^{m-|Y|}-1\right)$.
        \item If $\psi(\mathbf{x}_{3}\mid Y) = 0$ and $\psi(\mathbf{x}_{1}\mid W) = 1$, then $\wt(c_{D}^{(2)}(\mathbf{x}_{1},\mathbf{x}_{2},\mathbf{x}_{3},\mathbf{x}_{4})) = \frac{|D|}{2}.$ 
        
        In this case, $\#(\mathbf{x}_{1},\mathbf{x}_{2},\mathbf{x}_{3},\mathbf{x}_{4}) = \left(2^{m-|W|}-1 \right) \times \left((2^{|Y|}-1)2^{m-|Y|} \right)$.
        \item If $\psi(\mathbf{x}_{3}\mid Y) = 1$ and $\psi(\mathbf{x}_{1}\mid W) = 1$, then $\wt(c_{D}^{(2)}(\mathbf{x}_{1},\mathbf{x}_{2},\mathbf{x}_{3},\mathbf{x}_{4})) = (2^{m}-2^{|X|})(2^{m}-2^{|Z|})2^{m+|W|-1}.$ 
        
        In this case, $\#(\mathbf{x}_{1},\mathbf{x}_{2},\mathbf{x}_{3},\mathbf{x}_{4}) = \left(2^{m-|W|}-1 \right) \times \left( 2^{m-|Y|}-1 \right)$.
    \end{enumerate}
\item[(11)] 
$\boldsymbol{\mathbf{x}_{1}+\mathbf{x}_{4} = 0, \mathbf{x}_{3} = 0, \mathbf{x}_{2} \neq 0, \mathbf{x}_{1} \neq 0} \implies \mathbf{x}_{1} \neq 0, \mathbf{x}_{2} \neq 0, \mathbf{x}_{3} = 0, \mathbf{x}_{4} = \mathbf{x}_{1}.$ Then
\begin{equation*}
    \wt(c_{D}^{(2)}(\mathbf{x}_{1},\mathbf{x}_{2},\mathbf{x}_{3},\mathbf{x}_{4})) = \frac{|D|}{2}+\frac{1}{2}(2^{m}-2^{|X|})(2^{m}-2^{|Y|})2^{|Z|+|W|}\psi(\mathbf{x}_{2}\mid Z)\psi(\mathbf{x}_{1}\mid W)
\end{equation*}
    \begin{enumerate}
        \item If $\psi(\mathbf{x}_{2}\mid Z) = 0$ and $\psi(\mathbf{x}_{1}\mid W) = 0$, then $\wt(c_{D}^{(2)}(\mathbf{x}_{1},\mathbf{x}_{2},\mathbf{x}_{3},\mathbf{x}_{4})) = \frac{|D|}{2}.$ 
        
        In this case, $\#(\mathbf{x}_{1},\mathbf{x}_{2},\mathbf{x}_{3},\mathbf{x}_{4}) = \left((2^{|W|}-1)2^{m-|W|} \right) \times \left( (2^{|Z|}-1)2^{m-|Z|}\right)$.
        \item If $\psi(\mathbf{x}_{2}\mid Z) = 1$ and $\psi(\mathbf{x}_{1}\mid W) = 0$, then $\wt(c_{D}^{(2)}(\mathbf{x}_{1},\mathbf{x}_{2},\mathbf{x}_{3},\mathbf{x}_{4})) = \frac{|D|}{2}.$ 
        
        In this case, $\#(\mathbf{x}_{1},\mathbf{x}_{2},\mathbf{x}_{3},\mathbf{x}_{4}) =\left((2^{|W|}-1)2^{m-|W|} \right) \times \left(2^{m-|Z|}-1 \right)$.
        \item If $\psi(\mathbf{x}_{2}\mid Z) = 0$ and $\psi(\mathbf{x}_{1}\mid W) = 1$, then $\wt(c_{D}^{(2)}(\mathbf{x}_{1},\mathbf{x}_{2},\mathbf{x}_{3},\mathbf{x}_{4})) = \frac{|D|}{2}.$ 
        
        In this case, $\#(\mathbf{x}_{1},\mathbf{x}_{2},\mathbf{x}_{3},\mathbf{x}_{4}) =\left(2^{m-|W|}-1 \right) \times \left((2^{|Z|}-1)2^{m-|Z|} \right)$.
        \item If $\psi(\mathbf{x}_{2}\mid Z) = 1$ and $\psi(\mathbf{x}_{1}\mid W) = 1$, then $\wt(c_{D}^{(2)}(\mathbf{x}_{1},\mathbf{x}_{2},\mathbf{x}_{3},\mathbf{x}_{4})) = (2^{m}-2^{|X|})(2^{m}-2^{|Y|})2^{m+|W|-1}.$ 
        
        In this case, $\#(\mathbf{x}_{1},\mathbf{x}_{2},\mathbf{x}_{3},\mathbf{x}_{4}) =\left(2^{m-|W|}-1 \right) \times \left(2^{m-|Z|}-1 \right)$.
    \end{enumerate}
\item[(12)] 
$\boldsymbol{\mathbf{x}_{1}+\mathbf{x}_{4} \neq 0, \mathbf{x}_{3} \neq 0, \mathbf{x}_{2} \neq 0, \mathbf{x}_{1} = 0} \implies \mathbf{x}_{1} = 0, \mathbf{x}_{2} \neq 0, \mathbf{x}_{3} \neq 0, \mathbf{x}_{4} \neq 0.$ Then
\begin{equation*}
    \wt(c_{D}^{(2)}(\mathbf{x}_{1},\mathbf{x}_{2},\mathbf{x}_{3},\mathbf{x}_{4})) = \frac{|D|}{2}+\frac{1}{2}2^{|X|+|Y|+|Z|+|W|}\psi(\mathbf{x}_{4}\mid X)\psi(\mathbf{x}_{3} \mid Y)\psi(\mathbf{x}_{2}\mid Z)
\end{equation*}
    \begin{enumerate}
        \item If $\psi(\mathbf{x}_{4}\mid X) = 0, \psi(\mathbf{x}_{3}\mid Y) = 0$ and $\psi(\mathbf{x}_{2}\mid Z) = 0$, then $\wt(c_{D}^{(2)}(\mathbf{x}_{1},\mathbf{x}_{2},\mathbf{x}_{3},\mathbf{x}_{4})) = \frac{|D|}{2}.$ 
        
        In this case, $\#(\mathbf{x}_{1},\mathbf{x}_{2},\mathbf{x}_{3},\mathbf{x}_{4}) = \left((2^{|Z|}-1)2^{m-|Z|} \right) \times \left( (2^{|Y|}-1)2^{m-|Y|}\right) \times \left((2^{|X|}-1)2^{m-|X|} \right)$.
        \item If $\psi(\mathbf{x}_{4}\mid X) = 1, \psi(\mathbf{x}_{3}\mid Y) = 0$ and $\psi(\mathbf{x}_{2}\mid Z) = 0$, then $\wt(c_{D}^{(2)}(\mathbf{x}_{1},\mathbf{x}_{2},\mathbf{x}_{3},\mathbf{x}_{4})) = \frac{|D|}{2}.$ 
        
        In this case, $\#(\mathbf{x}_{1},\mathbf{x}_{2},\mathbf{x}_{3},\mathbf{x}_{4}) = \left((2^{|Z|}-1)2^{m-|Z|} \right) \times \left((2^{|Y|}-1)2^{m-|Y|} \right) \times \left( 2^{m-|X|}-1\right)$.
        \item If $\psi(\mathbf{x}_{4}\mid X) = 0, \psi(\mathbf{x}_{3}\mid Y) = 1$ and $\psi(\mathbf{x}_{2}\mid Z) = 0$, then $\wt(c_{D}^{(2)}(\mathbf{x}_{1},\mathbf{x}_{2},\mathbf{x}_{3},\mathbf{x}_{4})) = \frac{|D|}{2}.$ 
        
        In this case, $\#(\mathbf{x}_{1},\mathbf{x}_{2},\mathbf{x}_{3},\mathbf{x}_{4}) =\left((2^{|Z|}-1)2^{m-|Z|} \right) \times \left( 2^{m-|Y|}-1\right) \times \left((2^{|X|}-1)2^{m-|X|} \right) $.
        \item If $\psi(\mathbf{x}_{4}\mid X) = 0, \psi(\mathbf{x}_{3}\mid Y) = 0$ and $\psi(\mathbf{x}_{2}\mid Z) = 1$, then $\wt(c_{D}^{(2)}(\mathbf{x}_{1},\mathbf{x}_{2},\mathbf{x}_{3},\mathbf{x}_{4})) = \frac{|D|}{2}.$ 
        
        In this case, $\#(\mathbf{x}_{1},\mathbf{x}_{2},\mathbf{x}_{3},\mathbf{x}_{4}) = \left(2^{m-|Z|}-1 \right) \times \left( (2^{|Y|}-1)2^{m-|Y|}\right) \times \left((2^{|X|}-1)2^{m-|X|} \right) $.
        \item If $\psi(\mathbf{x}_{4}\mid X) = 1, \psi(\mathbf{x}_{3}\mid Y) = 1$ and $\psi(\mathbf{x}_{2}\mid Z) = 0$, then $\wt(c_{D}^{(2)}(\mathbf{x}_{1},\mathbf{x}_{2},\mathbf{x}_{3},\mathbf{x}_{4})) = \frac{|D|}{2}.$ 
        
        In this case, $\#(\mathbf{x}_{1},\mathbf{x}_{2},\mathbf{x}_{3},\mathbf{x}_{4}) = \left((2^{|Z|}-1)2^{m-|Z|} \right) \times \left(2^{m-|Y|}-1 \right) \times \left(2^{m-|X|}-1 \right) $.
        \item If $\psi(\mathbf{x}_{4}\mid X) = 1, \psi(\mathbf{x}_{3}\mid Y) = 0$ and $\psi(\mathbf{x}_{2}\mid Z) = 1$, then $\wt(c_{D}^{(2)}(\mathbf{x}_{1},\mathbf{x}_{2},\mathbf{x}_{3},\mathbf{x}_{4})) = \frac{|D|}{2}.$ 
        
        In this case, $\#(\mathbf{x}_{1},\mathbf{x}_{2},\mathbf{x}_{3},\mathbf{x}_{4}) = \left(2^{m-|Z|}-1 \right) \times \left( (2^{|Y|}-1)2^{m-|Y|}\right) \times \left( 2^{m-|X|}-1 \right) $.
        \item If $\psi(\mathbf{x}_{4}\mid X) = 0, \psi(\mathbf{x}_{3}\mid Y) = 1$ and $\psi(\mathbf{x}_{2}\mid Z) = 1$, then $\wt(c_{D}^{(2)}(\mathbf{x}_{1},\mathbf{x}_{2},\mathbf{x}_{3},\mathbf{x}_{4})) = \frac{|D|}{2}.$ 
        
        In this case, $\#(\mathbf{x}_{1},\mathbf{x}_{2},\mathbf{x}_{3},\mathbf{x}_{4}) = \left(2^{m-|Z|}-1 \right) \times \left( 2^{m-|Y|}-1 \right) \times \left((2^{|X|}-1)2^{m-|X|} \right) $.
        \item If $\psi(\mathbf{x}_{4}\mid X) = 1, \psi(\mathbf{x}_{3}\mid Y) = 1$ and $\psi(\mathbf{x}_{2}\mid Z) = 1$, then $\wt(c_{D}^{(2)}(\mathbf{x}_{1},\mathbf{x}_{2},\mathbf{x}_{3},\mathbf{x}_{4})) = (2^{2m}-2^{m+|X|}-2^{m+|Y|}-2^{m+|Z|}+2^{|X|+|Y|}+2^{|Y|+|Z|}+2^{|X|+|Z|})2^{m+|W|-1}.$ 
        
        In this case, $\#(\mathbf{x}_{1},\mathbf{x}_{2},\mathbf{x}_{3},\mathbf{x}_{4}) = \left(2^{m-|Z|}-1 \right) \times \left(2^{m-|Y|}-1 \right) \times \left( 2^{m-|X|}-1\right) $.
    \end{enumerate}
    \item[(13)] 
$\boldsymbol{\mathbf{x}_{1}+\mathbf{x}_{4} \neq 0, \mathbf{x}_{3} \neq 0, \mathbf{x}_{2} = 0, \mathbf{x}_{1} \neq 0}.$ Then
\begin{equation*}
    \wt(c_{D}^{(2)}(\mathbf{x}_{1},\mathbf{x}_{2},\mathbf{x}_{3},\mathbf{x}_{4})) = \frac{|D|}{2}-\frac{1}{2}(2^{m}-2^{|Z|})2^{|X|+|Y|+|W|}\psi(\mathbf{x}_{1}+\mathbf{x}_{4}\mid X)\psi(\mathbf{x}_{3} \mid Y)\psi(\mathbf{x}_{1}\mid W).
\end{equation*}
Subcase - (I) $\mathbf{x}_{4} = 0$

$\implies \mathbf{x}_{1} \neq 0, \mathbf{x}_{2} = 0, \mathbf{x}_{3} \neq 0, \mathbf{x}_{4} = 0$. Then
    \begin{equation*}
    \wt(c_{D}^{(2)}(\mathbf{x}_{1},\mathbf{x}_{2},\mathbf{x}_{3},\mathbf{x}_{4})) = \frac{|D|}{2}-\frac{1}{2}(2^{m}-2^{|Z|})2^{|X|+|Y|+|W|}\psi(\mathbf{x}_{1}\mid X)\psi(\mathbf{x}_{3} \mid Y)\psi(\mathbf{x}_{1}\mid W).
\end{equation*}
\begin{enumerate}
    \item If $\psi(\mathbf{x}_{1}\mid X) = 0, \psi(\mathbf{x}_{3}\mid Y) = 0$ and $\psi(\mathbf{x}_{1}\mid W) = 0$, then $\wt(c_{D}^{(2)}(\mathbf{x}_{1},\mathbf{x}_{2},\mathbf{x}_{3},\mathbf{x}_{4})) = \frac{|D|}{2}.$ 
        
        In this case, $\#(\mathbf{x}_{1},\mathbf{x}_{2},\mathbf{x}_{3},\mathbf{x}_{4}) = \left(2^{m}-2^{m-|X|}-2^{m-|W|}+2^{m-|X\cup W|} \right) \times \left((2^{|Y|}-1)2^{m-|Y|} \right)$.
        \item If $\psi(\mathbf{x}_{1}\mid X) = 1, \psi(\mathbf{x}_{3}\mid Y) = 0$ and $\psi(\mathbf{x}_{1}\mid W) = 0$, then $\wt(c_{D}^{(2)}(\mathbf{x}_{1},\mathbf{x}_{2},\mathbf{x}_{3},\mathbf{x}_{4})) = \frac{|D|}{2}.$ 
        
        In this case, $\#(\mathbf{x}_{1},\mathbf{x}_{2},\mathbf{x}_{3},\mathbf{x}_{4}) = \left((2^{|W\setminus X|}-1)2^{m-|X\cup W|} \right) \times \left( (2^{|Y|}-1)2^{m-|Y|}\right)$.
        \item If $\psi(\mathbf{x}_{1}\mid X) = 0, \psi(\mathbf{x}_{3}\mid Y) = 1$ and $\psi(\mathbf{x}_{1}\mid W) = 0$, then $\wt(c_{D}^{(2)}(\mathbf{x}_{1},\mathbf{x}_{2},\mathbf{x}_{3},\mathbf{x}_{4})) = \frac{|D|}{2}.$ 
        
        In this case, $\#(\mathbf{x}_{1},\mathbf{x}_{2},\mathbf{x}_{3},\mathbf{x}_{4}) = \left(2^{m}-2^{m-|X|}-2^{m-|W|}+2^{m-|X\cup W|}\right) \times \left(2^{m-|Y|}-1 \right) $.
        \item If $\psi(\mathbf{x}_{1}\mid X) = 0, \psi(\mathbf{x}_{3}\mid Y) = 0$ and $\psi(\mathbf{x}_{1}\mid W) = 1$, then $\wt(c_{D}^{(2)}(\mathbf{x}_{1},\mathbf{x}_{2},\mathbf{x}_{3},\mathbf{x}_{4})) = \frac{|D|}{2}.$ 
        
        In this case, $\#(\mathbf{x}_{1},\mathbf{x}_{2},\mathbf{x}_{3},\mathbf{x}_{4}) = \left((2^{|X\setminus W|}-1)2^{m-|X\cup W|} \right) \times \left((2^{|Y|}-1)2^{m-|Y|} \right) $.
        \item If $\psi(\mathbf{x}_{1}\mid X) = 1, \psi(\mathbf{x}_{3}\mid Y) = 1$ and $\psi(\mathbf{x}_{1}\mid W) = 0$, then $\wt(c_{D}^{(2)}(\mathbf{x}_{1},\mathbf{x}_{2},\mathbf{x}_{3},\mathbf{x}_{4})) = \frac{|D|}{2}.$ 
        
        In this case, $\#(\mathbf{x}_{1},\mathbf{x}_{2},\mathbf{x}_{3},\mathbf{x}_{4}) =\left((2^{|W\setminus X|}-1)2^{m-|X\cup W|} \right) \times \left(2^{m-|Y|}-1 \right) $.
        \item If $\psi(\mathbf{x}_{1}\mid X) = 1, \psi(\mathbf{x}_{3}\mid Y) = 0$ and $\psi(\mathbf{x}_{1}\mid W) = 1$, then $\wt(c_{D}^{(2)}(\mathbf{x}_{1},\mathbf{x}_{2},\mathbf{x}_{3},\mathbf{x}_{4})) = \frac{|D|}{2}.$ 
        
        In this case, $\#(\mathbf{x}_{1},\mathbf{x}_{2},\mathbf{x}_{3},\mathbf{x}_{4}) = \left(2^{m-|X\cup W|}-1 \right) \times \left((2^{|Y|}-1)2^{m-|Y|} \right) $.
        \item If $\psi(\mathbf{x}_{1}\mid X) = 0, \psi(\mathbf{x}_{3}\mid Y) = 1$ and $\psi(\mathbf{x}_{1}\mid W) = 1$, then $\wt(c_{D}^{(2)}(\mathbf{x}_{1},\mathbf{x}_{2},\mathbf{x}_{3},\mathbf{x}_{4})) = \frac{|D|}{2}.$ 
        
        In this case, $\#(\mathbf{x}_{1},\mathbf{x}_{2},\mathbf{x}_{3},\mathbf{x}_{4}) = \left((2^{|X\setminus W|}-1)2^{m-|X\cup W|} \right) \times \left(2^{m-|Y|}-1 \right) $.
        \item If $\psi(\mathbf{x}_{1}\mid X) = 1, \psi(\mathbf{x}_{3}\mid Y) = 1$ and $\psi(\mathbf{x}_{1}\mid W) = 1$, then $\wt(c_{D}^{(2)}(\mathbf{x}_{1},\mathbf{x}_{2},\mathbf{x}_{3},\mathbf{x}_{4})) = (2^{m}-2^{|Z|})(2^{m}-2^{|X|}-2^{|Y|})2^{m+|W|-1}.$ 
        
        In this case, $\#(\mathbf{x}_{1},\mathbf{x}_{2},\mathbf{x}_{3},\mathbf{x}_{4}) = \left(2^{m-|X\cup W|}-1 \right) \times \left(2^{m-|Y|}-1 \right) $.
\end{enumerate}
Subcase - (II) $\mathbf{x}_{4} \neq 0$

$\implies \mathbf{x}_{1} \neq 0, \mathbf{x}_{2} = 0, \mathbf{x}_{3} \neq 0, \mathbf{x}_{4} \neq 0, $ and $\mathbf{x}_{1} \neq \mathbf{x}_{4}$. Then
\begin{equation*}
    \wt(c_{D}^{(2)}(\mathbf{x}_{1},\mathbf{x}_{2},\mathbf{x}_{3},\mathbf{x}_{4})) = \frac{|D|}{2}-\frac{1}{2}(2^{m}-2^{|Z|})2^{|X|+|Y|+|W|}\psi(\mathbf{x}_{1}+\mathbf{x}_{4}\mid X)\psi(\mathbf{x}_{3} \mid Y)\psi(\mathbf{x}_{1}\mid W).
\end{equation*}
\begin{enumerate}
    \item If $\psi(\mathbf{x}_{1}+\mathbf{x}_{4}\mid X) = 0, \psi(\mathbf{x}_{3}\mid Y) = 0$ and $\psi(\mathbf{x}_{1}\mid W) = 0$, then $\wt(c_{D}^{(2)}(\mathbf{x}_{1},\mathbf{x}_{2},\mathbf{x}_{3},\mathbf{x}_{4})) = \frac{|D|}{2}.$
    \begin{enumerate}
        \item $\mathbf{x}_{4} \cap X = \emptyset$

        In this case, $\#(\mathbf{x}_{1},\mathbf{x}_{2},\mathbf{x}_{3},\mathbf{x}_{4}) = \left(2^{m}-2^{m-|X|}-2^{m-|W|}+2^{m-|X\cup W|} \right) \times \left((2^{|Y|}-1)2^{m-|Y|} \right) \times \left(2^{m-|X|}-1 \right)$.
        \item $\mathbf{x}_{4} \cap X \neq \emptyset$
        \begin{itemize}
            \item $\mathbf{x}_{1} \cap X = \emptyset$

            In this case,  $ \#(\mathbf{x}_{1},\mathbf{x}_{2},\mathbf{x}_{3},\mathbf{x}_{4}) = \left((2^{|W\setminus X|}-1)2^{m-|X\cup W
            |} \right) \times \left((2^{|Y|}-1)2^{m-|Y|} \right) \times \\ \left( (2^{|X|}-1)2^{m-|X|}\right) $.
            \item $\mathbf{x}_{1} \cap X \neq \emptyset$

            In this case,  $\#(\mathbf{x}_{1},\mathbf{x}_{2},\mathbf{x}_{3},\mathbf{x}_{4}) =\left(2^{m}-2^{m-|X|}-2^{m-|W|}+2^{m-|X\cup W|}\right)\\ \times \left((2^{|Y|}-1)2^{m-|Y|} \right) \times \left((2^{|X|}-2)2^{m-|X|} \right)$.
        \end{itemize}
    \end{enumerate}
     \item If $\psi(\mathbf{x}_{1}+\mathbf{x}_{4}\mid X) = 1, \psi(\mathbf{x}_{3}\mid Y) = 0$ and $\psi(\mathbf{x}_{1}\mid W) = 0$, then $\wt(c_{D}^{(2)}(\mathbf{x}_{1},\mathbf{x}_{2},\mathbf{x}_{3},\mathbf{x}_{4})) = \frac{|D|}{2}.$
     \begin{enumerate}
            \item $\mathbf{x}_{4} \cap X = \emptyset$

            In this case, $\#(\mathbf{x}_{1},\mathbf{x}_{2},\mathbf{x}_{3},\mathbf{x}_{4}) =\left( (2^{|W\setminus X|}-1)2^{m-|X\cup W|} \right) \times \left((2^{|Y|}-1)2^{m-|Y|} \right) \times \left( 2^{m-|X|}-2 \right)$.
            \item $\mathbf{x}_{4} \cap X \neq \emptyset $

             In this case, $\#(\mathbf{x}_{1},\mathbf{x}_{2},\mathbf{x}_{3},\mathbf{x}_{4}) =\left(2^{m}-2^{m-|X|}-2^{m-|W|}+2^{m-|X\cup W|}\right) \times \left((2^{|Y|}-1)2^{m-|Y|} \right) \times \left(2^{m-|X|}-1 \right) $.
        \end{enumerate}
        \item  If $\psi(\mathbf{x}_{1}+\mathbf{x}_{4}\mid X) = 0, \psi(\mathbf{x}_{3}\mid Y) = 1$ and $\psi(\mathbf{x}_{1}\mid W) = 0$, then $\wt(c_{D}^{(2)}(\mathbf{x}_{1},\mathbf{x}_{2},\mathbf{x}_{3},\mathbf{x}_{4})) = \frac{|D|}{2}.$
        \begin{enumerate}
    \item $\mathbf{x}_{4} \cap X = \emptyset$

In this case, $\#(\mathbf{x}_{1},\mathbf{x}_{2},\mathbf{x}_{3},\mathbf{x}_{4}) =\left(2^{m}-2^{m-|X|}-2^{m-|W|}+2^{m-|X\cup W|}\right) \times \left(  2^{m-|Y|}-1\right) \times \left( 2^{m-|X|}-1 \right)$.
\item $\mathbf{x}_{4} \cap X \neq \emptyset$
\begin{itemize}
    \item $\mathbf{x}_{1} \cap X = \emptyset$

In this case, $\#(\mathbf{x}_{1},\mathbf{x}_{2},\mathbf{x}_{3},\mathbf{x}_{4}) =\left((2^{|W\setminus X|}-1)2^{m-|X\cup W|} \right) \times \left( 2^{m-|Y|}-1\right) \times \\ \left((2^{|X|}-1)2^{m-|X|} \right)$. 
\item $\mathbf{x}_{1} \cap X \neq \emptyset$

In this case, $\#(\mathbf{x}_{1},\mathbf{x}_{2},\mathbf{x}_{3},\mathbf{x}_{4}) =\left(2^{m}-2^{m-|X|}-2^{m-|W|}+2^{m-|X\cup W|}\right) \times \left(2^{m-|Y|}-1 \right) \times \left((2^{|X|}-2)2^{m-|X|} \right)$.
\end{itemize}
\end{enumerate}
 \item  If $\psi(\mathbf{x}_{1}+\mathbf{x}_{4}\mid X) = 0, \psi(\mathbf{x}_{3}\mid Y) = 0$ and $\psi(\mathbf{x}_{1}\mid W) = 1$, then $\wt(c_{D}^{(2)}(\mathbf{x}_{1},\mathbf{x}_{2},\mathbf{x}_{3},\mathbf{x}_{4})) = \frac{|D|}{2}.$
 \begin{enumerate}
    \item $\mathbf{x}_{4} \cap X = \emptyset$

In this case, $\#(\mathbf{x}_{1},\mathbf{x}_{2},\mathbf{x}_{3},\mathbf{x}_{4}) =\left((2^{|X\setminus W|}-1)2^{m-|X\cup W|} \right) \times \left((2^{|Y|}-1)2^{m-|Y|} \right) \times \left(2^{m-|X|}-1 \right)$.
\item $\mathbf{x}_{4} \cap X \neq \emptyset$
\begin{itemize}
    \item $\mathbf{x}_{1} \cap X = \emptyset$

In this case, $\#(\mathbf{x}_{1},\mathbf{x}_{2},\mathbf{x}_{3},\mathbf{x}_{4}) =\left(2^{m-|X\cup W|}-1 \right) \times \left((2^{|Y|}-1)2^{m-|Y|} \right) \times \\ \left((2^{|X|}-1)2^{m-|X|} \right) $. 
\item $\mathbf{x}_{1} \cap X \neq \emptyset$

In this case, $\#(\mathbf{x}_{1},\mathbf{x}_{2},\mathbf{x}_{3},\mathbf{x}_{4}) =\left( (2^{|X\setminus W|}-1)2^{m-|X\cup W|}\right) \times \left( (2^{|Y|}-1)2^{m-|Y|}\right) \times \\ \left((2^{|X|}-2)2^{m-|X|} \right)$.
\end{itemize}
\end{enumerate}
 \item If $\psi(\mathbf{x}_{1}+\mathbf{x}_{4}\mid X) = 1, \psi(\mathbf{x}_{3}\mid Y) = 1$ and $\psi(\mathbf{x}_{1}\mid W) = 0$, then $\wt(c_{D}^{(2)}(\mathbf{x}_{1},\mathbf{x}_{2},\mathbf{x}_{3},\mathbf{x}_{4})) = \frac{|D|}{2}.$
  \begin{enumerate}
            \item $\mathbf{x}_{4} \cap X = \emptyset$

            In this case, $\#(\mathbf{x}_{1},\mathbf{x}_{2},\mathbf{x}_{3},\mathbf{x}_{4}) =\left( (2^{|W\setminus X|}-1)2^{m-|X\cup W|}\right) \times \left( 2^{m-|Y|}-1\right) \times \left(2^{m-|X|}-2 \right) $.
            \item $\mathbf{x}_{4} \cap X \neq \emptyset $

             In this case, $\#(\mathbf{x}_{1},\mathbf{x}_{2},\mathbf{x}_{3},\mathbf{x}_{4}) =\left( 2^{m}-2^{m-|X|}-2^{m-|W|}+2^{m-|X\cup W|}\right) \times \left(2^{m-|Y|}-1 \right) \times \left(2^{m-|X|}-1 \right) $.
        \end{enumerate}
         \item If $\psi(\mathbf{x}_{1}+\mathbf{x}_{4}\mid X) = 1, \psi(\mathbf{x}_{3}\mid Y) = 0$ and $\psi(\mathbf{x}_{1}\mid W) = 1$, then $\wt(c_{D}^{(2)}(\mathbf{x}_{1},\mathbf{x}_{2},\mathbf{x}_{3},\mathbf{x}_{4})) = \frac{|D|}{2}.$
         \begin{enumerate}
            \item $\mathbf{x}_{4} \cap X = \emptyset$

            In this case, $\#(\mathbf{x}_{1},\mathbf{x}_{2},\mathbf{x}_{3},\mathbf{x}_{4}) =\left(2^{m-|X\cup W|}-1 \right) \times \left((2^{|Y|}-1)2^{m-|Y|} \right) \times \left(2^{m-|X|}-2 \right) $.
            \item $\mathbf{x}_{4} \cap X \neq \emptyset $

             In this case, $\#(\mathbf{x}_{1},\mathbf{x}_{2},\mathbf{x}_{3},\mathbf{x}_{4}) =\left((2^{|X\setminus W|}-1)2^{m-|X\cup W|} \right) \times \left((2^{|Y|}-1)2^{m-|Y|} \right) \times \left(2^{m-|X|}-1 \right) $.
        \end{enumerate}
        \item If $\psi(\mathbf{x}_{1}+\mathbf{x}_{4}\mid X) = 0, \psi(\mathbf{x}_{3}\mid Y) = 1$ and $\psi(\mathbf{x}_{1}\mid W) = 1$, then $\wt(c_{D}^{(2)}(\mathbf{x}_{1},\mathbf{x}_{2},\mathbf{x}_{3},\mathbf{x}_{4})) = \frac{|D|}{2}.$
         \begin{enumerate}
    \item $\mathbf{x}_{4} \cap X = \emptyset$

In this case, $\#(\mathbf{x}_{1},\mathbf{x}_{2},\mathbf{x}_{3},\mathbf{x}_{4}) =\left((2^{|X\setminus W|}-1)2^{m-|X\cup W|} \right) \times \left(2^{m-|Y|}-1 \right) \times \left(2^{m-|X|}-1 \right)$.
\item $\mathbf{x}_{4} \cap X \neq \emptyset$
\begin{itemize}
    \item $\mathbf{x}_{1} \cap X = \emptyset$

In this case, $\#(\mathbf{x}_{1},\mathbf{x}_{2},\mathbf{x}_{3},\mathbf{x}_{4}) =\left( 2^{m-|X\cup W|}-1\right) \times \left(2^{m-|Y|}-1 \right) \times \left((2^{|X|}-1)2^{m-|X|} \right)$. 
\item $\mathbf{x}_{1} \cap X \neq \emptyset$

In this case, $\#(\mathbf{x}_{1},\mathbf{x}_{2},\mathbf{x}_{3},\mathbf{x}_{4}) =\left((2^{|X\setminus W|}-1)2^{m-|X\cup W|} \right) \times \left(2^{m-|Y|}-1 \right) \times \\ \left((2^{|X|}-2)2^{m-|X|} \right) $.
\end{itemize}
\end{enumerate}
        \item If $\psi(\mathbf{x}_{1}+\mathbf{x}_{4}\mid X) = 1, \psi(\mathbf{x}_{3}\mid Y) = 1$ and $\psi(\mathbf{x}_{1}\mid W) = 1$, then $\wt(c_{D}^{(2)}(\mathbf{x}_{1},\mathbf{x}_{2},\mathbf{x}_{3},\mathbf{x}_{4})) = (2^{m}-2^{|Z|})(2^{m}-2^{|X|}-2^{|Y|})2^{m+|W|-1}.$
        \begin{enumerate}
            \item $\mathbf{x}_{4} \cap X = \emptyset$

            In this case, $\#(\mathbf{x}_{1},\mathbf{x}_{2},\mathbf{x}_{3},\mathbf{x}_{4}) =\left(2^{m-|X\cup W|}-1 \right) \times \left(2^{m-|Y|}-1 \right) \times \left(2^{m-|X|}-2 \right) $.
            \item $\mathbf{x}_{4} \cap X \neq \emptyset $

             In this case, $\#(\mathbf{x}_{1},\mathbf{x}_{2},\mathbf{x}_{3},\mathbf{x}_{4}) =\left((2^{|X\setminus W|}-1)2^{m-|X\cup W|} \right) \times \left(2^{m-|Y|}-1 \right) \times \left(2^{m-|X|}-1 \right)$.
        \end{enumerate}
\end{enumerate}
\item[(14)] 
$\boldsymbol{\mathbf{x}_{1}+\mathbf{x}_{4} \neq 0, \mathbf{x}_{3} = 0, \mathbf{x}_{2} \neq 0, \mathbf{x}_{1} \neq 0}.$ Then
\begin{equation*}
    \wt(c_{D}^{(2)}(\mathbf{x}_{1},\mathbf{x}_{2},\mathbf{x}_{3},\mathbf{x}_{4})) = \frac{|D|}{2}-\frac{1}{2}(2^{m}-2^{|Y|})2^{|X|+|Z|+|W|}\psi(\mathbf{x}_{1}+\mathbf{x}_{4}\mid X)\psi(\mathbf{x}_{2} \mid Z)\psi(\mathbf{x}_{1}\mid W).
\end{equation*}
Subcase - (I) $\mathbf{x}_{4} = 0$

$\implies \mathbf{x}_{1} \neq 0, \mathbf{x}_{2} \neq 0, \mathbf{x}_{3} = 0, \mathbf{x}_{4} = 0$. Then
    \begin{equation*}
    \wt(c_{D}^{(2)}(\mathbf{x}_{1},\mathbf{x}_{2},\mathbf{x}_{3},\mathbf{x}_{4})) = \frac{|D|}{2}-\frac{1}{2}(2^{m}-2^{|Y|})2^{|X|+|Z|+|W|}\psi(\mathbf{x}_{1}\mid X)\psi(\mathbf{x}_{2} \mid Z)\psi(\mathbf{x}_{1}\mid W).
\end{equation*}
\begin{enumerate}
    \item If $\psi(\mathbf{x}_{1}\mid X) = 0, \psi(\mathbf{x}_{2}\mid Z) = 0$ and $\psi(\mathbf{x}_{1}\mid W) = 0$, then $\wt(c_{D}^{(2)}(\mathbf{x}_{1},\mathbf{x}_{2},\mathbf{x}_{3},\mathbf{x}_{4})) = \frac{|D|}{2}.$ 
        
        In this case, $\#(\mathbf{x}_{1},\mathbf{x}_{2},\mathbf{x}_{3},\mathbf{x}_{4}) = \left(2^{m}-2^{m-|X|}-2^{m-|W|}+2^{m-|X\cup W|} \right) \times \left((2^{|Z|}-1)2^{m-|Z|} \right)$.
        \item If $\psi(\mathbf{x}_{1}\mid X) = 1, \psi(\mathbf{x}_{2}\mid Z) = 0$ and $\psi(\mathbf{x}_{1}\mid W) = 0$, then $\wt(c_{D}^{(2)}(\mathbf{x}_{1},\mathbf{x}_{2},\mathbf{x}_{3},\mathbf{x}_{4})) = \frac{|D|}{2}.$ 
        
        In this case, $\#(\mathbf{x}_{1},\mathbf{x}_{2},\mathbf{x}_{3},\mathbf{x}_{4}) = \left((2^{|W\setminus X|}-1)2^{m-|X\cup W|} \right) \times \left((2^{|Z|}-1)2^{m-|Z|} \right) $.
        \item If $\psi(\mathbf{x}_{1}\mid X) = 0, \psi(\mathbf{x}_{2}\mid Z) = 1$ and $\psi(\mathbf{x}_{1}\mid W) = 0$, then $\wt(c_{D}^{(2)}(\mathbf{x}_{1},\mathbf{x}_{2},\mathbf{x}_{3},\mathbf{x}_{4})) = \frac{|D|}{2}.$ 
        
        In this case, $\#(\mathbf{x}_{1},\mathbf{x}_{2},\mathbf{x}_{3},\mathbf{x}_{4}) = \left(2^{m}-2^{m-|X|}-2^{m-|W|}+2^{m-|X\cup W|} \right) \times \left(2^{m-|Z|}-1 \right) $.
        \item If $\psi(\mathbf{x}_{1}\mid X) = 0, \psi(\mathbf{x}_{2}\mid Z) = 0$ and $\psi(\mathbf{x}_{1}\mid W) = 1$, then $\wt(c_{D}^{(2)}(\mathbf{x}_{1},\mathbf{x}_{2},\mathbf{x}_{3},\mathbf{x}_{4})) = \frac{|D|}{2}.$ 
        
        In this case, $\#(\mathbf{x}_{1},\mathbf{x}_{2},\mathbf{x}_{3},\mathbf{x}_{4}) =\left((2^{|X\setminus W|}-1)2^{m-|X\cup W|} \right) \times \left( (2^{|Z|}-1)2^{m-|Z|} \right) $.
        \item If $\psi(\mathbf{x}_{1}\mid X) = 1, \psi(\mathbf{x}_{2}\mid Z) = 1$ and $\psi(\mathbf{x}_{1}\mid W) = 0$, then $\wt(c_{D}^{(2)}(\mathbf{x}_{1},\mathbf{x}_{2},\mathbf{x}_{3},\mathbf{x}_{4})) = \frac{|D|}{2}.$ 
        
        In this case, $\#(\mathbf{x}_{1},\mathbf{x}_{2},\mathbf{x}_{3},\mathbf{x}_{4}) = \left((2^{|W\setminus X|}-1)2^{m-|X\cup W|} \right) \times \left(2^{m-|Z|}-1 \right) $.
        \item If $\psi(\mathbf{x}_{1}\mid X) = 1, \psi(\mathbf{x}_{2}\mid Z) = 0$ and $\psi(\mathbf{x}_{1}\mid W) = 1$, then $\wt(c_{D}^{(2)}(\mathbf{x}_{1},\mathbf{x}_{2},\mathbf{x}_{3},\mathbf{x}_{4})) = \frac{|D|}{2}.$ 
        
        In this case, $\#(\mathbf{x}_{1},\mathbf{x}_{2},\mathbf{x}_{3},\mathbf{x}_{4}) = \left(2^{m-|X\cup W|}-1 \right) \times \left((2^{|Z|}-1)2^{m-|Z|} \right) $.
        \item If $\psi(\mathbf{x}_{1}\mid X) = 0, \psi(\mathbf{x}_{2}\mid Z) = 1$ and $\psi(\mathbf{x}_{1}\mid W) = 1$, then $\wt(c_{D}^{(2)}(\mathbf{x}_{1},\mathbf{x}_{2},\mathbf{x}_{3},\mathbf{x}_{4})) = \frac{|D|}{2}.$ 
        
        In this case, $\#(\mathbf{x}_{1},\mathbf{x}_{2},\mathbf{x}_{3},\mathbf{x}_{4}) = \left((2^{|X\setminus W|}-1)2^{m-|X\cup W|} \right) \times \left(2^{m-|Z|}-1 \right)$.
        \item If $\psi(\mathbf{x}_{1}\mid X) = 1, \psi(\mathbf{x}_{2}\mid Z) = 1$ and $\psi(\mathbf{x}_{1}\mid W) = 1$, then $\wt(c_{D}^{(2)}(\mathbf{x}_{1},\mathbf{x}_{2},\mathbf{x}_{3},\mathbf{x}_{4})) = (2^{m}-2^{|Y|})(2^{m}-2^{|X|}-2^{|Z|})2^{m+|W|-1}.$ 
        
        In this case, $\#(\mathbf{x}_{1},\mathbf{x}_{2},\mathbf{x}_{3},\mathbf{x}_{4}) = \left(2^{m-|X\cup W|}-1 \right) \times \left(2^{m-|Z|}-1 \right) $.
\end{enumerate}
Subcase - (II) $\mathbf{x}_{4} \neq 0$

$\implies \mathbf{x}_{1} \neq 0, \mathbf{x}_{2} \neq 0, \mathbf{x}_{3} = 0, \mathbf{x}_{4} \neq 0, $ and $\mathbf{x}_{1} \neq \mathbf{x}_{4}$. Then
\begin{equation*}
    \wt(c_{D}^{(2)}(\mathbf{x}_{1},\mathbf{x}_{2},\mathbf{x}_{3},\mathbf{x}_{4})) = \frac{|D|}{2}-\frac{1}{2}(2^{m}-2^{|Y|})2^{|X|+|Z|+|W|}\psi(\mathbf{x}_{1}+\mathbf{x}_{4}\mid X)\psi(\mathbf{x}_{2} \mid Z)\psi(\mathbf{x}_{1}\mid W).
\end{equation*}
\begin{enumerate}
    \item If $\psi(\mathbf{x}_{1}+\mathbf{x}_{4}\mid X) = 0, \psi(\mathbf{x}_{2}\mid Z) = 0$ and $\psi(\mathbf{x}_{1}\mid W) = 0$, then $\wt(c_{D}^{(2)}(\mathbf{x}_{1},\mathbf{x}_{2},\mathbf{x}_{3},\mathbf{x}_{4})) = \frac{|D|}{2}.$
    \begin{enumerate}
        \item $\mathbf{x}_{4} \cap X = \emptyset$

        In this case, $\#(\mathbf{x}_{1},\mathbf{x}_{2},\mathbf{x}_{3},\mathbf{x}_{4}) =\left(2^{m}-2^{m-|X|}-2^{m-|W|}+2^{m-|X\cup W|} \right) \times \\ \left((2^{|Z|}-1)2^{m-|Z|} \right) \times \left(2^{m-|X|}-1 \right)$.
        \item $\mathbf{x}_{4} \cap X \neq \emptyset$
        \begin{itemize}
            \item $\mathbf{x}_{1} \cap X = \emptyset$

            In this case,  $\#(\mathbf{x}_{1},\mathbf{x}_{2},\mathbf{x}_{3},\mathbf{x}_{4}) =\left((2^{|W\setminus X|}-1)2^{m-|X\cup W
            |} \right) \times \left((2^{|Z|}-1)2^{m-|Z|} \right) \times \\ \left((2^{|X|}-1)2^{m-|X|} \right) $.
            \item $\mathbf{x}_{1} \cap X \neq \emptyset$

            In this case,  $\#(\mathbf{x}_{1},\mathbf{x}_{2},\mathbf{x}_{3},\mathbf{x}_{4}) =\left(2^{m}-2^{m-|X|}-2^{m-|W|}+2^{m-|X\cup W|}\right) \times \\ \left((2^{|Z|}-1)2^{m-|Z|} \right) \times \left((2^{|X|}-2)2^{m-|X|} \right)$.
        \end{itemize}
    \end{enumerate}
     \item If $\psi(\mathbf{x}_{1}+\mathbf{x}_{4}\mid X) = 1, \psi(\mathbf{x}_{2}\mid Z) = 0$ and $\psi(\mathbf{x}_{1}\mid W) = 0$, then $\wt(c_{D}^{(2)}(\mathbf{x}_{1},\mathbf{x}_{2},\mathbf{x}_{3},\mathbf{x}_{4})) = \frac{|D|}{2}.$
     \begin{enumerate}
            \item $\mathbf{x}_{4} \cap X = \emptyset$

            In this case, $\#(\mathbf{x}_{1},\mathbf{x}_{2},\mathbf{x}_{3},\mathbf{x}_{4}) =\left((2^{|W\setminus X|}-1)2^{m-|X\cup W|} \right) \times \left((2^{|Z|}-1)2^{m-|Z|} \right) \times \\ \left(2^{m-|X|}-2 \right)$.
            \item $\mathbf{x}_{4} \cap X \neq \emptyset $

             In this case, $\#(\mathbf{x}_{1},\mathbf{x}_{2},\mathbf{x}_{3},\mathbf{x}_{4}) =\left(2^{m}-2^{m-|X|}-2^{m-|W|}+2^{m-|X\cup W|}\right) \times \\ \left( (2^{|Z|}-1)2^{m-|Z|}\right) \times \left( 2^{m-|X|}-1 \right)$.
        \end{enumerate}
        \item  If $\psi(\mathbf{x}_{1}+\mathbf{x}_{4}\mid X) = 0, \psi(\mathbf{x}_{2}\mid Z) = 1$ and $\psi(\mathbf{x}_{1}\mid W) = 0$, then $\wt(c_{D}^{(2)}(\mathbf{x}_{1},\mathbf{x}_{2},\mathbf{x}_{3},\mathbf{x}_{4})) = \frac{|D|}{2}.$
        \begin{enumerate}
    \item $\mathbf{x}_{4} \cap X = \emptyset$

In this case, $\#(\mathbf{x}_{1},\mathbf{x}_{2},\mathbf{x}_{3},\mathbf{x}_{4}) =\left(2^{m}-2^{m-|X|}-2^{m-|W|}+2^{m-|X\cup W|}\right) \times \left( 2^{m-|Z|}-1\right) \times \left( 2^{m-|X|}-1\right)$.
\item $\mathbf{x}_{4} \cap X \neq \emptyset$
\begin{itemize}
    \item $\mathbf{x}_{1} \cap X = \emptyset$

In this case, $\#(\mathbf{x}_{1},\mathbf{x}_{2},\mathbf{x}_{3},\mathbf{x}_{4}) =\left((2^{|W\setminus X|}-1)2^{m-|X\cup W|} \right) \times \left(2^{m-|Z|}-1 \right) \times \\ \left( (2^{|X|}-1)2^{m-|X|}\right)$. 
\item $\mathbf{x}_{1} \cap X \neq \emptyset$

In this case, $\#(\mathbf{x}_{1},\mathbf{x}_{2},\mathbf{x}_{3},\mathbf{x}_{4}) =\left(2^{m}-2^{m-|X|}-2^{m-|W|}+2^{m-|X\cup W|}\right) \times \left(2^{m-|Z|}-1 \right) \times \left((2^{|X|}-2)2^{m-|X|} \right)$.
\end{itemize}
\end{enumerate}
 \item  If $\psi(\mathbf{x}_{1}+\mathbf{x}_{4}\mid X) = 0, \psi(\mathbf{x}_{2}\mid Z) = 0$ and $\psi(\mathbf{x}_{1}\mid W) = 1$, then $\wt(c_{D}^{(2)}(\mathbf{x}_{1},\mathbf{x}_{2},\mathbf{x}_{3},\mathbf{x}_{4})) = \frac{|D|}{2}.$
 \begin{enumerate}
    \item $\mathbf{x}_{4} \cap X = \emptyset$

In this case, $\#(\mathbf{x}_{1},\mathbf{x}_{2},\mathbf{x}_{3},\mathbf{x}_{4}) =\left((2^{|X\setminus W|}-1)2^{m-|X\cup W|} \right) \times \left((2^{|Z|}-1)2^{m-|Z|} \right) \times \\ \left(2^{m-|X|}-1 \right)$.
\item $\mathbf{x}_{4} \cap X \neq \emptyset$
\begin{itemize}
    \item $\mathbf{x}_{1} \cap X = \emptyset$

In this case, $\#(\mathbf{x}_{1},\mathbf{x}_{2},\mathbf{x}_{3},\mathbf{x}_{4}) =\left(2^{m-|X\cup W|}-1 \right) \times \left((2^{|Z|}-1)2^{m-|Z|} \right) \times \\ \left((2^{|X|}-1)2^{m-|X|} \right)$. 
\item $\mathbf{x}_{1} \cap X \neq \emptyset$

In this case, $\#(\mathbf{x}_{1},\mathbf{x}_{2},\mathbf{x}_{3},\mathbf{x}_{4}) =\left((2^{|X\setminus W|}-1)2^{m-|X\cup W|} \right) \times \left((2^{|Z|}-1)2^{m-|Z|} \right) \times \\  \left((2^{|X|}-2)2^{m-|X|} \right)$.
\end{itemize}
\end{enumerate}
 \item If $\psi(\mathbf{x}_{1}+\mathbf{x}_{4}\mid X) = 1, \psi(\mathbf{x}_{2}\mid Z) = 1$ and $\psi(\mathbf{x}_{1}\mid W) = 0$, then $\wt(c_{D}^{(2)}(\mathbf{x}_{1},\mathbf{x}_{2},\mathbf{x}_{3},\mathbf{x}_{4})) = \frac{|D|}{2}.$
  \begin{enumerate}
            \item $\mathbf{x}_{4} \cap X = \emptyset$

            In this case, $\#(\mathbf{x}_{1},\mathbf{x}_{2},\mathbf{x}_{3},\mathbf{x}_{4}) =\left((2^{|W\setminus X|}-1)2^{m-|X\cup W|} \right) \times \left( 2^{m-|Z|}-1 \right) \times \left(2^{m-|X|}-2 \right) $.
            \item $\mathbf{x}_{4} \cap X \neq \emptyset $

             In this case, $\#(\mathbf{x}_{1},\mathbf{x}_{2},\mathbf{x}_{3},\mathbf{x}_{4}) =\left(2^{m}-2^{m-|X|}-2^{m-|W|}+2^{m-|X\cup W|}\right) \times \left(2^{m-|Z|}-1 \right) \times \left(2^{m-|X|}-1 \right)$.
        \end{enumerate}
         \item If $\psi(\mathbf{x}_{1}+\mathbf{x}_{4}\mid X) = 1, \psi(\mathbf{x}_{2}\mid Z) = 0$ and $\psi(\mathbf{x}_{1}\mid W) = 1$, then $\wt(c_{D}^{(2)}(\mathbf{x}_{1},\mathbf{x}_{2},\mathbf{x}_{3},\mathbf{x}_{4})) = \frac{|D|}{2}.$
         \begin{enumerate}
            \item $\mathbf{x}_{4} \cap X = \emptyset$

            In this case, $\#(\mathbf{x}_{1},\mathbf{x}_{2},\mathbf{x}_{3},\mathbf{x}_{4}) =\left( 2^{m-|X\cup W|}-1\right) \times \left((2^{|Z|}-1)2^{m-|Z|} \right) \times \left( 2^{m-|X|}-2\right)$.
            \item $\mathbf{x}_{4} \cap X \neq \emptyset $

             In this case, $\#(\mathbf{x}_{1},\mathbf{x}_{2},\mathbf{x}_{3},\mathbf{x}_{4}) =\left((2^{|X\setminus W|}-1)2^{m-|X\cup W|} \right) \times \left((2^{|Z|}-1)2^{m-|Z|} \right) \times \\ \left(2^{m-|X|}-1 \right) $.
        \end{enumerate}
        \item If $\psi(\mathbf{x}_{1}+\mathbf{x}_{4}\mid X) = 0, \psi(\mathbf{x}_{2}\mid Z) = 1$ and $\psi(\mathbf{x}_{1}\mid W) = 1$, then $\wt(c_{D}^{(2)}(\mathbf{x}_{1},\mathbf{x}_{2},\mathbf{x}_{3},\mathbf{x}_{4})) = \frac{|D|}{2}.$
         \begin{enumerate}
    \item $\mathbf{x}_{4} \cap X = \emptyset$

In this case, $\#(\mathbf{x}_{1},\mathbf{x}_{2},\mathbf{x}_{3},\mathbf{x}_{4}) =\left((2^{|X\setminus W|}-1)2^{m-|X\cup W|} \right) \times \left(2^{m-|Z|}-1 \right) \times \left( 2^{m-|X|}-1\right)$.
\item $\mathbf{x}_{4} \cap X \neq \emptyset$
\begin{itemize}
    \item $\mathbf{x}_{1} \cap X = \emptyset$

In this case, $\#(\mathbf{x}_{1},\mathbf{x}_{2},\mathbf{x}_{3},\mathbf{x}_{4}) =\left(2^{m-|X\cup W|}-1 \right) \times \left(2^{m-|Z|}-1 \right) \times \left((2^{|X|}-1)2^{m-|X|} \right) $. 
\item $\mathbf{x}_{1} \cap X \neq \emptyset$

In this case, $\#(\mathbf{x}_{1},\mathbf{x}_{2},\mathbf{x}_{3},\mathbf{x}_{4}) =\left((2^{|X\setminus W|}-1)2^{m-|X\cup W|} \right) \times \left(2^{m-|Z|}-1 \right) \times \\  \left((2^{|X|}-2)2^{m-|X|} \right)$.
\end{itemize}
\end{enumerate}
        \item If $\psi(\mathbf{x}_{1}+\mathbf{x}_{4}\mid X) = 1, \psi(\mathbf{x}_{2}\mid Z) = 1$ and $\psi(\mathbf{x}_{1}\mid W) = 1$, then $\wt(c_{D}^{(2)}(\mathbf{x}_{1},\mathbf{x}_{2},\mathbf{x}_{3},\mathbf{x}_{4})) = (2^{m}-2^{|Y|})(2^{m}-2^{|X|}-2^{|Z|})2^{m+|W|-1}.$
        \begin{enumerate}
            \item $\mathbf{x}_{4} \cap X = \emptyset$

            In this case, $\#(\mathbf{x}_{1},\mathbf{x}_{2},\mathbf{x}_{3},\mathbf{x}_{4}) =\left(2^{m-|X\cup W|}-1 \right) \times \left( 2^{m-|Z|}-1 \right) \times \left(2^{m-|X|}-2 \right)$.
            \item $\mathbf{x}_{4} \cap X \neq \emptyset $

             In this case, $\#(\mathbf{x}_{1},\mathbf{x}_{2},\mathbf{x}_{3},\mathbf{x}_{4}) =\left((2^{|X\setminus W|}-1)2^{m-|X\cup W|} \right) \times \left(2^{m-|Z|}-1 \right) \times \left( 2^{m-|X|}-1\right)$.
        \end{enumerate}
        \end{enumerate}
        \item[(15)] 
$\boldsymbol{\mathbf{x}_{1}+\mathbf{x}_{4} = 0, \mathbf{x}_{3} \neq 0, \mathbf{x}_{2} \neq 0, \mathbf{x}_{1} \neq 0} \implies \mathbf{x}_{1}\neq 0, \mathbf{x}_{2} \neq 0, \mathbf{x}_{3} \neq 0, \mathbf{x}_{4} = \mathbf{x}_{1}.$ Then
\begin{equation*}
    \wt(c_{D}^{(2)}(\mathbf{x}_{1},\mathbf{x}_{2},\mathbf{x}_{3},\mathbf{x}_{4})) = \frac{|D|}{2}-\frac{1}{2}(2^{m}-2^{|X|})2^{|Y|+|Z|+|W|}\psi(\mathbf{x}_{3} \mid Y)\psi(\mathbf{x}_{2}\mid Z)\psi(\mathbf{x}_{1}\mid W).
\end{equation*}
\begin{enumerate}
    \item If $\psi(\mathbf{x}_{3}\mid Y) = 0, \psi(\mathbf{x}_{2}\mid Z) = 0$ and $\psi(\mathbf{x}_{1}\mid W) = 0$, then $\wt(c_{D}^{(2)}(\mathbf{x}_{1},\mathbf{x}_{2},\mathbf{x}_{3},\mathbf{x}_{4})) = \frac{|D|}{2}.$ 
        
         In this case, $\#(\mathbf{x}_{1},\mathbf{x}_{2},\mathbf{x}_{3},\mathbf{x}_{4}) =\left( (2^{|W|}-1)2^{m-|W|}\right) \times \left((2^{|Z|}-1)2^{m-|Z|} \right) \times \left((2^{|Y|}-1)2^{m-|Y|} \right)$.
        \item If $\psi(\mathbf{x}_{3}\mid Y) = 1, \psi(\mathbf{x}_{2}\mid Z) = 0$ and $\psi(\mathbf{x}_{1}\mid W) = 0$, then $\wt(c_{D}^{(2)}(\mathbf{x}_{1},\mathbf{x}_{2},\mathbf{x}_{3},\mathbf{x}_{4})) = \frac{|D|}{2}.$ 
        
        In this case, $\#(\mathbf{x}_{1},\mathbf{x}_{2},\mathbf{x}_{3},\mathbf{x}_{4}) = \left((2^{|W|}-1)2^{m-|W|} \right) \times \left((2^{|Z|}-1)2^{m-|Z|} \right) \times \left(2^{m-|Y|}-1 \right) $.
        \item If $\psi(\mathbf{x}_{3}\mid Y) = 0, \psi(\mathbf{x}_{2}\mid Z) = 1$ and $\psi(\mathbf{x}_{1}\mid W) = 0$, then $\wt(c_{D}^{(2)}(\mathbf{x}_{1},\mathbf{x}_{2},\mathbf{x}_{3},\mathbf{x}_{4})) = \frac{|D|}{2}.$ 
        
        In this case, $\#(\mathbf{x}_{1},\mathbf{x}_{2},\mathbf{x}_{3},\mathbf{x}_{4}) = \left((2^{|W|}-1)2^{m-|W|} \right) \times \left(2^{m-|Z|}-1 \right) \times \left((2^{|Y|}-1)2^{m-|Y|} \right) $.
        \item If $\psi(\mathbf{x}_{3}\mid Y) = 0, \psi(\mathbf{x}_{2}\mid Z) = 0$ and $\psi(\mathbf{x}_{1}\mid W) = 1$, then $\wt(c_{D}^{(2)}(\mathbf{x}_{1},\mathbf{x}_{2},\mathbf{x}_{3},\mathbf{x}_{4})) = \frac{|D|}{2}.$ 
        
        In this case, $\#(\mathbf{x}_{1},\mathbf{x}_{2},\mathbf{x}_{3},\mathbf{x}_{4}) = \left( 2^{m-|W|}-1\right) \times \left((2^{|Z|}-1)2^{m-|Z|} \right) \times \left( (2^{|Y|}-1)2^{m-|Y|}\right) $.
        \item If $\psi(\mathbf{x}_{3}\mid Y) = 1, \psi(\mathbf{x}_{2}\mid Z) = 1$ and $\psi(\mathbf{x}_{1}\mid W) = 0$, then $\wt(c_{D}^{(2)}(\mathbf{x}_{1},\mathbf{x}_{2},\mathbf{x}_{3},\mathbf{x}_{4})) = \frac{|D|}{2}.$ 
        
       In this case, $\#(\mathbf{x}_{1},\mathbf{x}_{2},\mathbf{x}_{3},\mathbf{x}_{4}) = \left((2^{|W|}-1)2^{m-|W|} \right) \times \left(2^{m-|Z|}-1 \right) \times \left(2^{m-|Y|}-1 \right) $.
        \item If $\psi(\mathbf{x}_{3}\mid Y) = 1, \psi(\mathbf{x}_{2}\mid Z) = 0$ and $\psi(\mathbf{x}_{1}\mid W) = 1$, then $\wt(c_{D}^{(2)}(\mathbf{x}_{1},\mathbf{x}_{2},\mathbf{x}_{3},\mathbf{x}_{4})) = \frac{|D|}{2}.$ 
        
        In this case, $\#(\mathbf{x}_{1},\mathbf{x}_{2},\mathbf{x}_{3},\mathbf{x}_{4}) = \left(2^{m-|W|}-1 \right) \times \left((2^{|Z|}-1)2^{m-|Z|} \right) \times \left(2^{m-|Y|}-1 \right) $.
        \item If $\psi(\mathbf{x}_{3}\mid Y) = 0, \psi(\mathbf{x}_{2}\mid Z) = 1$ and $\psi(\mathbf{x}_{1}\mid W) = 1$, then $\wt(c_{D}^{(2)}(\mathbf{x}_{1},\mathbf{x}_{2},\mathbf{x}_{3},\mathbf{x}_{4})) = \frac{|D|}{2}.$ 
        
        In this case, $\#(\mathbf{x}_{1},\mathbf{x}_{2},\mathbf{x}_{3},\mathbf{x}_{4}) = \left(2^{m-|W|}-1 \right) \times \left(2^{m-|Z|}-1 \right) \times \left((2^{|Y|}-1)2^{m-|Y|} \right) $.
        \item If $\psi(\mathbf{x}_{3}\mid Y) = 1, \psi(\mathbf{x}_{2}\mid Z) = 1$ and $\psi(\mathbf{x}_{1}\mid W) = 1$, then $\wt(c_{D}^{(2)}(\mathbf{x}_{1},\mathbf{x}_{2},\mathbf{x}_{3},\mathbf{x}_{4})) = (2^{m}-2^{|X|})(2^{m}-2^{|Y|}-2^{|Z|})2^{m+|W|-1}.$ 
        
        In this case, $\#(\mathbf{x}_{1},\mathbf{x}_{2},\mathbf{x}_{3},\mathbf{x}_{4}) = \left( 2^{m-|W|}-1\right) \times \left( 2^{m-|Z|}-1\right) \times \left(2^{m-|Y|}-1 \right) $.
\end{enumerate}
\item[(16)] 
$\boldsymbol{\mathbf{x}_{1}+\mathbf{x}_{4} \neq 0, \mathbf{x}_{3} \neq 0, \mathbf{x}_{2} \neq 0, \mathbf{x}_{1} \neq 0}$. Then
\begin{equation*}
    \wt(c_{D}^{(2)}(\mathbf{x}_{1},\mathbf{x}_{2},\mathbf{x}_{3},\mathbf{x}_{4})) = \frac{|D|}{2}+\frac{1}{2} 2^{|X|+|Y|+|Z|+|W|}\psi(\mathbf{x}_{1}+\mathbf{x}_{4} \mid X)\psi(\mathbf{x}_{3} \mid Y)\psi(\mathbf{x}_{2}\mid Z)\psi(\mathbf{x}_{1}\mid W).
\end{equation*}
Subcase - (I) $\mathbf{x}_{4} = 0$

$\implies \mathbf{x}_{1} \neq 0, \mathbf{x}_{2} \neq 0, \mathbf{x}_{3} \neq 0, \mathbf{x}_{4} = 0.
$ Then
\begin{equation*}
    \wt(c_{D}^{(2)}(\mathbf{x}_{1},\mathbf{x}_{2},\mathbf{x}_{3},\mathbf{x}_{4})) = \frac{|D|}{2}+\frac{1}{2} 2^{|X|+|Y|+|Z|+|W|}\psi(\mathbf{x}_{1} \mid X)\psi(\mathbf{x}_{3} \mid Y)\psi(\mathbf{x}_{2}\mid Z)\psi(\mathbf{x}_{1}\mid W).
\end{equation*}
\begin{enumerate}
    \item If $\psi(\mathbf{x}_{1}\mid X) = 0, \psi(\mathbf{x}_{3}\mid Y) = 0, \psi(\mathbf{x}_{2}\mid Z) = 0$ and $\psi(\mathbf{x}_{1}\mid W) = 0$, then $\wt(c_{D}^{(2)}(\mathbf{x}_{1},\mathbf{x}_{2},\mathbf{x}_{3},\mathbf{x}_{4})) = \frac{|D|}{2}$.

    In this case, $\#(\mathbf{x}_{1},\mathbf{x}_{2},\mathbf{x}_{3},\mathbf{x}_{4}) = (13, \text{ Subcase - I, (a)})\times (2^{m}-2^{m-|Z|})$.
    \item If $\psi(\mathbf{x}_{1}\mid X) = 1, \psi(\mathbf{x}_{3}\mid Y) = 0, \psi(\mathbf{x}_{2}\mid Z) = 0$ and $\psi(\mathbf{x}_{1}\mid W) = 0$, then $\wt(c_{D}^{(2)}(\mathbf{x}_{1},\mathbf{x}_{2},\mathbf{x}_{3},\mathbf{x}_{4})) = \frac{|D|}{2}$.

    In this case, $\#(\mathbf{x}_{1},\mathbf{x}_{2},\mathbf{x}_{3},\mathbf{x}_{4}) = (13, \text{ Subcase - I, (b)})\times (2^{m}-2^{m-|Z|})$.
    \item If $\psi(\mathbf{x}_{1}\mid X) = 0, \psi(\mathbf{x}_{3}\mid Y) = 1, \psi(\mathbf{x}_{2}\mid Z) = 0$ and $\psi(\mathbf{x}_{1}\mid W) = 0$, then $\wt(c_{D}^{(2)}(\mathbf{x}_{1},\mathbf{x}_{2},\mathbf{x}_{3},\mathbf{x}_{4})) = \frac{|D|}{2}$.

    In this case, $\#(\mathbf{x}_{1},\mathbf{x}_{2},\mathbf{x}_{3},\mathbf{x}_{4}) = (13, \text{ Subcase - I, (c)})\times (2^{m}-2^{m-|Z|})$.
    \item If $\psi(\mathbf{x}_{1}\mid X) = 0, \psi(\mathbf{x}_{3}\mid Y) = 0, \psi(\mathbf{x}_{2}\mid Z) = 1$ and $\psi(\mathbf{x}_{1}\mid W) = 0$, then $\wt(c_{D}^{(2)}(\mathbf{x}_{1},\mathbf{x}_{2},\mathbf{x}_{3},\mathbf{x}_{4})) = \frac{|D|}{2}$.

    In this case, $\#(\mathbf{x}_{1},\mathbf{x}_{2},\mathbf{x}_{3},\mathbf{x}_{4}) = (13, \text{ Subcase - I, (a)})\times (2^{m-|Z|}-1)$.
    \item If $\psi(\mathbf{x}_{1}\mid X) = 0, \psi(\mathbf{x}_{3}\mid Y) = 0, \psi(\mathbf{x}_{2}\mid Z) = 0$ and $\psi(\mathbf{x}_{1}\mid W) = 1$, then $\wt(c_{D}^{(2)}(\mathbf{x}_{1},\mathbf{x}_{2},\mathbf{x}_{3},\mathbf{x}_{4})) = \frac{|D|}{2}$.

    In this case, $\#(\mathbf{x}_{1},\mathbf{x}_{2},\mathbf{x}_{3},\mathbf{x}_{4}) = (13, \text{ Subcase - I, (d)})\times (2^{m}-2^{m-|Z|})$.
    \item If $\psi(\mathbf{x}_{1}\mid X) = 1, \psi(\mathbf{x}_{3}\mid Y) = 1, \psi(\mathbf{x}_{2}\mid Z) = 0$ and $\psi(\mathbf{x}_{1}\mid W) = 0$, then $\wt(c_{D}^{(2)}(\mathbf{x}_{1},\mathbf{x}_{2},\mathbf{x}_{3},\mathbf{x}_{4})) = \frac{|D|}{2}$.

    In this case, $\#(\mathbf{x}_{1},\mathbf{x}_{2},\mathbf{x}_{3},\mathbf{x}_{4}) = (13, \text{ Subcase - I, (e)})\times (2^{m}-2^{m-|Z|})$.
    \item If $\psi(\mathbf{x}_{1}\mid X) = 0, \psi(\mathbf{x}_{3}\mid Y) = 1, \psi(\mathbf{x}_{2}\mid Z) = 1$ and $\psi(\mathbf{x}_{1}\mid W) = 0$, then $\wt(c_{D}^{(2)}(\mathbf{x}_{1},\mathbf{x}_{2},\mathbf{x}_{3},\mathbf{x}_{4})) = \frac{|D|}{2}$.

    In this case, $\#(\mathbf{x}_{1},\mathbf{x}_{2},\mathbf{x}_{3},\mathbf{x}_{4}) = (13, \text{ Subcase - I, (c)})\times (2^{m-|Z|}-1)$.
    \item If $\psi(\mathbf{x}_{1}\mid X) = 0, \psi(\mathbf{x}_{3}\mid Y) = 0, \psi(\mathbf{x}_{2}\mid Z) = 1$ and $\psi(\mathbf{x}_{1}\mid W) = 1$, then $\wt(c_{D}^{(2)}(\mathbf{x}_{1},\mathbf{x}_{2},\mathbf{x}_{3},\mathbf{x}_{4})) = \frac{|D|}{2}$.

    In this case, $\#(\mathbf{x}_{1},\mathbf{x}_{2},\mathbf{x}_{3},\mathbf{x}_{4}) = (13, \text{ Subcase - I, (d)})\times (2^{m-|Z|}-1)$.
    \item If $\psi(\mathbf{x}_{1}\mid X) = 1, \psi(\mathbf{x}_{3}\mid Y) = 0, \psi(\mathbf{x}_{2}\mid Z) = 0$ and $\psi(\mathbf{x}_{1}\mid W) = 1$, then $\wt(c_{D}^{(2)}(\mathbf{x}_{1},\mathbf{x}_{2},\mathbf{x}_{3},\mathbf{x}_{4})) = \frac{|D|}{2}$.

    In this case, $\#(\mathbf{x}_{1},\mathbf{x}_{2},\mathbf{x}_{3},\mathbf{x}_{4}) = (13, \text{ Subcase - I, (f)})\times (2^{m}-2^{m-|Z|})$.
    \item If $\psi(\mathbf{x}_{1}\mid X) = 1, \psi(\mathbf{x}_{3}\mid Y) = 0, \psi(\mathbf{x}_{2}\mid Z) = 1$ and $\psi(\mathbf{x}_{1}\mid W) = 0$, then $\wt(c_{D}^{(2)}(\mathbf{x}_{1},\mathbf{x}_{2},\mathbf{x}_{3},\mathbf{x}_{4})) = \frac{|D|}{2}$.

    In this case, $\#(\mathbf{x}_{1},\mathbf{x}_{2},\mathbf{x}_{3},\mathbf{x}_{4}) = (13, \text{ Subcase - I, (b)})\times (2^{m-|Z|}-1)$.
    \item If $\psi(\mathbf{x}_{1}\mid X) = 0, \psi(\mathbf{x}_{3}\mid Y) = 1, \psi(\mathbf{x}_{2}\mid Z) = 0$ and $\psi(\mathbf{x}_{1}\mid W) = 1$, then $\wt(c_{D}^{(2)}(\mathbf{x}_{1},\mathbf{x}_{2},\mathbf{x}_{3},\mathbf{x}_{4})) = \frac{|D|}{2}$.

    In this case, $\#(\mathbf{x}_{1},\mathbf{x}_{2},\mathbf{x}_{3},\mathbf{x}_{4}) = (13, \text{ Subcase - I, (g)})\times (2^{m}-2^{m-|Z|})$.
    \item If $\psi(\mathbf{x}_{1}\mid X) = 1, \psi(\mathbf{x}_{3}\mid Y) = 1, \psi(\mathbf{x}_{2}\mid Z) = 1$ and $\psi(\mathbf{x}_{1}\mid W) = 0$, then $\wt(c_{D}^{(2)}(\mathbf{x}_{1},\mathbf{x}_{2},\mathbf{x}_{3},\mathbf{x}_{4})) = \frac{|D|}{2}$.

    In this case, $\#(\mathbf{x}_{1},\mathbf{x}_{2},\mathbf{x}_{3},\mathbf{x}_{4}) = (13, \text{ Subcase - I, (e)})\times (2^{m-|Z|}-1)$.
    \item If $\psi(\mathbf{x}_{1}\mid X) = 0, \psi(\mathbf{x}_{3}\mid Y) = 1, \psi(\mathbf{x}_{2}\mid Z) = 1$ and $\psi(\mathbf{x}_{1}\mid W) = 1$, then $\wt(c_{D}^{(2)}(\mathbf{x}_{1},\mathbf{x}_{2},\mathbf{x}_{3},\mathbf{x}_{4})) = \frac{|D|}{2}$.

    In this case, $\#(\mathbf{x}_{1},\mathbf{x}_{2},\mathbf{x}_{3},\mathbf{x}_{4}) = (13, \text{ Subcase - I, (g)})\times (2^{m-|Z|}-1)$.
    \item If $\psi(\mathbf{x}_{1}\mid X) = 1, \psi(\mathbf{x}_{3}\mid Y) = 0, \psi(\mathbf{x}_{2}\mid Z) = 1$ and $\psi(\mathbf{x}_{1}\mid W) = 1$, then $\wt(c_{D}^{(2)}(\mathbf{x}_{1},\mathbf{x}_{2},\mathbf{x}_{3},\mathbf{x}_{4})) = \frac{|D|}{2}$.

    In this case, $\#(\mathbf{x}_{1},\mathbf{x}_{2},\mathbf{x}_{3},\mathbf{x}_{4}) = (13, \text{ Subcase - I, (f)})\times (2^{m-|Z|}-1)$.
    \item If $\psi(\mathbf{x}_{1}\mid X) = 1, \psi(\mathbf{x}_{3}\mid Y) = 1, \psi(\mathbf{x}_{2}\mid Z) = 0$ and $\psi(\mathbf{x}_{1}\mid W) = 1$, then $\wt(c_{D}^{(2)}(\mathbf{x}_{1},\mathbf{x}_{2},\mathbf{x}_{3},\mathbf{x}_{4})) = \frac{|D|}{2}$.

    In this case, $\#(\mathbf{x}_{1},\mathbf{x}_{2},\mathbf{x}_{3},\mathbf{x}_{4}) = (13, \text{ Subcase - I, (h)})\times (2^{m}-2^{m-|Z|})$.
    \item If $\psi(\mathbf{x}_{1}\mid X) = 1, \psi(\mathbf{x}_{3}\mid Y) = 1, \psi(\mathbf{x}_{2}\mid Z) = 1$ and $\psi(\mathbf{x}_{1}\mid W) = 1$, then $\wt(c_{D}^{(2)}(\mathbf{x}_{1},\mathbf{x}_{2},\mathbf{x}_{3},\mathbf{x}_{4})) = (2^{2m}-2^{m+|X|}-2^{m+|Y|}-2^{m+|Z|}+2^{|X|+|Y|}+2^{|Y|+|Z|}+2^{|X|+|Z|})2^{m+|W|-1}$.

    In this case, $\#(\mathbf{x}_{1},\mathbf{x}_{2},\mathbf{x}_{3},\mathbf{x}_{4}) = (13, \text{ Subcase - I, (h)})\times (2^{m-|Z|}-1)$.
\end{enumerate}
Subcase - (II) $\mathbf{x}_{4} \neq 0$

$\implies \mathbf{x}_{1} \neq 0, \mathbf{x}_{2} \neq 0, \mathbf{x}_{3} \neq 0, \mathbf{x}_{4} \neq 0,
$ and $\mathbf{x}_{1}\neq\mathbf{x}_{4}$. Then
\begin{equation*}
    \wt(c_{D}^{(2)}(\mathbf{x}_{1},\mathbf{x}_{2},\mathbf{x}_{3},\mathbf{x}_{4})) = \frac{|D|}{2}+\frac{1}{2} 2^{|X|+|Y|+|Z|+|W|}\psi(\mathbf{x}_{1}+\mathbf{x}_{4} \mid X)\psi(\mathbf{x}_{3} \mid Y)\psi(\mathbf{x}_{2}\mid Z)\psi(\mathbf{x}_{1}\mid W).
\end{equation*}
\begin{enumerate}
    \item If $\psi(\mathbf{x}_{1}+\mathbf{x}_{4}\mid X) = 0, \psi(\mathbf{x}_{3}\mid Y) = 0, \psi(\mathbf{x}_{2}\mid Z) = 0$ and $\psi(\mathbf{x}_{1}\mid W) = 0$, then $\wt(c_{D}^{(2)}(\mathbf{x}_{1},\mathbf{x}_{2},\mathbf{x}_{3},\mathbf{x}_{4})) = \frac{|D|}{2}$.

    In this case, $\#(\mathbf{x}_{1},\mathbf{x}_{2},\mathbf{x}_{3},\mathbf{x}_{4}) = (13, \text{ Subcase - II, (a)})\times (2^{m}-2^{m-|Z|})$.
    \item If $\psi(\mathbf{x}_{1}+\mathbf{x}_{4}\mid X) = 1, \psi(\mathbf{x}_{3}\mid Y) = 0, \psi(\mathbf{x}_{2}\mid Z) = 0$ and $\psi(\mathbf{x}_{1}\mid W) = 0$, then $\wt(c_{D}^{(2)}(\mathbf{x}_{1},\mathbf{x}_{2},\mathbf{x}_{3},\mathbf{x}_{4})) = \frac{|D|}{2}$.

    In this case, $\#(\mathbf{x}_{1},\mathbf{x}_{2},\mathbf{x}_{3},\mathbf{x}_{4}) = (13, \text{ Subcase - II, (b)})\times (2^{m}-2^{m-|Z|})$.
    \item If $\psi(\mathbf{x}_{1}+\mathbf{x}_{4}\mid X) = 0, \psi(\mathbf{x}_{3}\mid Y) = 1, \psi(\mathbf{x}_{2}\mid Z) = 0$ and $\psi(\mathbf{x}_{1}\mid W) = 0$, then $\wt(c_{D}^{(2)}(\mathbf{x}_{1},\mathbf{x}_{2},\mathbf{x}_{3},\mathbf{x}_{4})) = \frac{|D|}{2}$.

    In this case, $\#(\mathbf{x}_{1},\mathbf{x}_{2},\mathbf{x}_{3},\mathbf{x}_{4}) = (13, \text{ Subcase - II, (c)})\times (2^{m}-2^{m-|Z|})$.
    \item If $\psi(\mathbf{x}_{1}+\mathbf{x}_{4}\mid X) = 0, \psi(\mathbf{x}_{3}\mid Y) = 0, \psi(\mathbf{x}_{2}\mid Z) = 1$ and $\psi(\mathbf{x}_{1}\mid W) = 0$, then $\wt(c_{D}^{(2)}(\mathbf{x}_{1},\mathbf{x}_{2},\mathbf{x}_{3},\mathbf{x}_{4})) = \frac{|D|}{2}$.

    In this case, $\#(\mathbf{x}_{1},\mathbf{x}_{2},\mathbf{x}_{3},\mathbf{x}_{4}) = (13, \text{ Subcase - II, (a)})\times (2^{m-|Z|}-1)$.
    \item If $\psi(\mathbf{x}_{1}+\mathbf{x}_{4}\mid X) = 0, \psi(\mathbf{x}_{3}\mid Y) = 0, \psi(\mathbf{x}_{2}\mid Z) = 0$ and $\psi(\mathbf{x}_{1}\mid W) = 1$, then $\wt(c_{D}^{(2)}(\mathbf{x}_{1},\mathbf{x}_{2},\mathbf{x}_{3},\mathbf{x}_{4})) = \frac{|D|}{2}$.

    In this case, $\#(\mathbf{x}_{1},\mathbf{x}_{2},\mathbf{x}_{3},\mathbf{x}_{4}) = (13, \text{ Subcase - II, (d)})\times (2^{m}-2^{m-|Z|})$.
    \item If $\psi(\mathbf{x}_{1}+\mathbf{x}_{4}\mid X) = 1, \psi(\mathbf{x}_{3}\mid Y) = 1, \psi(\mathbf{x}_{2}\mid Z) = 0$ and $\psi(\mathbf{x}_{1}\mid W) = 0$, then $\wt(c_{D}^{(2)}(\mathbf{x}_{1},\mathbf{x}_{2},\mathbf{x}_{3},\mathbf{x}_{4})) = \frac{|D|}{2}$.

    In this case, $\#(\mathbf{x}_{1},\mathbf{x}_{2},\mathbf{x}_{3},\mathbf{x}_{4}) = (13, \text{ Subcase - II, (e)})\times (2^{m}-2^{m-|Z|})$.
    \item If $\psi(\mathbf{x}_{1}+\mathbf{x}_{4}\mid X) = 0, \psi(\mathbf{x}_{3}\mid Y) = 1, \psi(\mathbf{x}_{2}\mid Z) = 1$ and $\psi(\mathbf{x}_{1}\mid W) = 0$, then $\wt(c_{D}^{(2)}(\mathbf{x}_{1},\mathbf{x}_{2},\mathbf{x}_{3},\mathbf{x}_{4})) = \frac{|D|}{2}$.

    In this case, $\#(\mathbf{x}_{1},\mathbf{x}_{2},\mathbf{x}_{3},\mathbf{x}_{4}) = (13, \text{ Subcase - II, (c)})\times (2^{m-|Z|}-1)$.
    \item If $\psi(\mathbf{x}_{1}+\mathbf{x}_{4}\mid X) = 0, \psi(\mathbf{x}_{3}\mid Y) = 0, \psi(\mathbf{x}_{2}\mid Z) = 1$ and $\psi(\mathbf{x}_{1}\mid W) = 1$, then $\wt(c_{D}^{(2)}(\mathbf{x}_{1},\mathbf{x}_{2},\mathbf{x}_{3},\mathbf{x}_{4})) = \frac{|D|}{2}$.

    In this case, $\#(\mathbf{x}_{1},\mathbf{x}_{2},\mathbf{x}_{3},\mathbf{x}_{4}) = (13, \text{ Subcase - II, (d)})\times (2^{m-|Z|}-1)$.
    \item If $\psi(\mathbf{x}_{1}+\mathbf{x}_{4}\mid X) = 1, \psi(\mathbf{x}_{3}\mid Y) = 0, \psi(\mathbf{x}_{2}\mid Z) = 0$ and $\psi(\mathbf{x}_{1}\mid W) = 1$, then $\wt(c_{D}^{(2)}(\mathbf{x}_{1},\mathbf{x}_{2},\mathbf{x}_{3},\mathbf{x}_{4})) = \frac{|D|}{2}$.

    In this case, $\#(\mathbf{x}_{1},\mathbf{x}_{2},\mathbf{x}_{3},\mathbf{x}_{4}) = (13, \text{ Subcase - II, (f)})\times (2^{m}-2^{m-|Z|})$.
    \item If $\psi(\mathbf{x}_{1}+\mathbf{x}_{4}\mid X) = 1, \psi(\mathbf{x}_{3}\mid Y) = 0, \psi(\mathbf{x}_{2}\mid Z) = 1$ and $\psi(\mathbf{x}_{1}\mid W) = 0$, then $\wt(c_{D}^{(2)}(\mathbf{x}_{1},\mathbf{x}_{2},\mathbf{x}_{3},\mathbf{x}_{4})) = \frac{|D|}{2}$.

    In this case, $\#(\mathbf{x}_{1},\mathbf{x}_{2},\mathbf{x}_{3},\mathbf{x}_{4}) = (13, \text{ Subcase - II, (b)})\times (2^{m-|Z|}-1)$.
    \item If $\psi(\mathbf{x}_{1}+\mathbf{x}_{4}\mid X) = 0, \psi(\mathbf{x}_{3}\mid Y) = 1, \psi(\mathbf{x}_{2}\mid Z) = 0$ and $\psi(\mathbf{x}_{1}\mid W) = 1$, then $\wt(c_{D}^{(2)}(\mathbf{x}_{1},\mathbf{x}_{2},\mathbf{x}_{3},\mathbf{x}_{4})) = \frac{|D|}{2}$.

    In this case, $\#(\mathbf{x}_{1},\mathbf{x}_{2},\mathbf{x}_{3},\mathbf{x}_{4}) = (13, \text{ Subcase - II, (g)})\times (2^{m}-2^{m-|Z|})$.
    \item If $\psi(\mathbf{x}_{1}+\mathbf{x}_{4}\mid X) = 1, \psi(\mathbf{x}_{3}\mid Y) = 1, \psi(\mathbf{x}_{2}\mid Z) = 1$ and $\psi(\mathbf{x}_{1}\mid W) = 0$, then $\wt(c_{D}^{(2)}(\mathbf{x}_{1},\mathbf{x}_{2},\mathbf{x}_{3},\mathbf{x}_{4})) = \frac{|D|}{2}$.

    In this case, $\#(\mathbf{x}_{1},\mathbf{x}_{2},\mathbf{x}_{3},\mathbf{x}_{4}) = (13, \text{ Subcase - II, (e)})\times (2^{m-|Z|}-1)$.
    \item If $\psi(\mathbf{x}_{1}+\mathbf{x}_{4}\mid X) = 0, \psi(\mathbf{x}_{3}\mid Y) = 1, \psi(\mathbf{x}_{2}\mid Z) = 1$ and $\psi(\mathbf{x}_{1}\mid W) = 1$, then $\wt(c_{D}^{(2)}(\mathbf{x}_{1},\mathbf{x}_{2},\mathbf{x}_{3},\mathbf{x}_{4})) = \frac{|D|}{2}$.

    In this case, $\#(\mathbf{x}_{1},\mathbf{x}_{2},\mathbf{x}_{3},\mathbf{x}_{4}) = (13, \text{ Subcase - II, (g)})\times (2^{m-|Z|}-1)$.
    \item If $\psi(\mathbf{x}_{1}+\mathbf{x}_{4}\mid X) = 1, \psi(\mathbf{x}_{3}\mid Y) = 0, \psi(\mathbf{x}_{2}\mid Z) = 1$ and $\psi(\mathbf{x}_{1}\mid W) = 1$, then $\wt(c_{D}^{(2)}(\mathbf{x}_{1},\mathbf{x}_{2},\mathbf{x}_{3},\mathbf{x}_{4})) = \frac{|D|}{2}$.

    In this case, $\#(\mathbf{x}_{1},\mathbf{x}_{2},\mathbf{x}_{3},\mathbf{x}_{4}) = (13, \text{ Subcase - II, (f)})\times (2^{m-|Z|}-1)$.
    \item If $\psi(\mathbf{x}_{1}+\mathbf{x}_{4}\mid X) = 1, \psi(\mathbf{x}_{3}\mid Y) = 1, \psi(\mathbf{x}_{2}\mid Z) = 0$ and $\psi(\mathbf{x}_{1}\mid W) = 1$, then $\wt(c_{D}^{(2)}(\mathbf{x}_{1},\mathbf{x}_{2},\mathbf{x}_{3},\mathbf{x}_{4})) = \frac{|D|}{2}$.

    In this case, $\#(\mathbf{x}_{1},\mathbf{x}_{2},\mathbf{x}_{3},\mathbf{x}_{4}) = (13, \text{ Subcase - II, (h)})\times (2^{m}-2^{m-|Z|})$.
    \item If $\psi(\mathbf{x}_{1}+\mathbf{x}_{4}\mid X) = 1, \psi(\mathbf{x}_{3}\mid Y) = 1, \psi(\mathbf{x}_{2}\mid Z) = 1$ and $\psi(\mathbf{x}_{1}\mid W) = 1$, then $\wt(c_{D}^{(2)}(\mathbf{x}_{1},\mathbf{x}_{2},\mathbf{x}_{3},\mathbf{x}_{4})) = (2^{2m}-2^{m+|X|}-2^{m+|Y|}-2^{m+|Z|}+2^{|X|+|Y|}+2^{|Y|+|Z|}+2^{|X|+|Z|})2^{m+|W|-1}$.

    In this case, $\#(\mathbf{x}_{1},\mathbf{x}_{2},\mathbf{x}_{3},\mathbf{x}_{4}) = (13, \text{ Subcase - II, (h)})\times (2^{m-|Z|}-1)$.
\end{enumerate}
\end{enumerate}
This concludes the calculations for part $(4)$. Next, we establish the distance-optimality in part $(3)$. 
Here, $n=(2^{m}-2^{|X|})(2^{m}-2^{|Y|})2^{|Z|+|W|}, k = 2m+|Z|+|W|$ and $d = (2^{m}-2^{|X|}-2^{|Y|})2^{m+|Z|+|W|-1}$. Let $c=|Z|+|W|$. Consider 
\begin{eqnarray*}
    \sum\limits_{i=0}^{k-1} \left\lceil \frac{d+1}{2^{i}} \right\rceil &=& \sum\limits_{i=0}^{2m+c-1} \left\lceil \frac{2^{2m+c-1}-2^{m+|X|+c-1} -2^{m+|Y|+c-1}+1}{2^{i}} \right\rceil \\
    &=& \sum\limits_{i=0}^{2m+c-1} 2^{2m+c-1 - i} + \sum\limits_{i=0}^{2m+c-1} \left\lceil \frac{-(2^{m+c+|X|-1}+2^{m+c+|Y|-1}-1)}{2^{i}} \right\rceil \\
    &=& 2^{2m+c}-1 - \sum\limits_{i=0}^{2m+c-1} \left\lfloor\frac{2^{m+c+|X|-1}+2^{m+c+|Y|-1}-1}{2^{i}}  \right\rfloor.
\end{eqnarray*}
If $|X| > |Y|$, then we have
\begin{align*}
    \sum\limits_{i=0}^{k-1} \left\lceil \frac{d+1}{2^{i}} \right\rceil &= 2^{2m+c}-1 - \sum\limits_{i=0}^{m+c+|Y|-1} \left\lfloor\frac{2^{m+c+|X|-1}+2^{m+c+|Y|-1}-1}{2^{i}}  \right\rfloor\\ & - \sum\limits_{i=m+c+|Y|}^{m+c+|X|-1} \left\lfloor\frac{2^{m+c+|X|-1}+2^{m+c+|Y|-1}-1}{2^{i}}  \right\rfloor\\ & - \sum\limits_{m+c+|X|}^{2m+c-1} \left\lfloor\frac{2^{m+c+|X|-1}+2^{m+c+|Y|-1}-1}{2^{i}}  \right\rfloor\\
    &= 2^{2m+c}-1 - \left\{ (2^{m+c+|X|-1}+2^{m+c+|Y|-1}-1)+(2^{m+c+|X|-2}+2^{m+c+|Y|-2}-1) \right.\\ & \left. +\cdots +(2^{|X|-|Y|}+1-1) \right\}-\left\{ 2^{|X|-|Y|-1}+2^{|X|-|Y|-2}+\cdots +1 \right\} \\
    &= \left( 2^{2m+c}-2^{m+c+|X|}-2^{m+c+|Y|} \right)+m+c+|Y|+1 \\
    &= (n-2^{|X|+|Y|+|Z|+|W|})+m+|Z|+|W|+|Y|+1 \\
    &\geq n+1 \\
    &\iff -2^{|X|+|Y|+|Z|+|W|} +m+|Z|+|W|+|Y| \geq 0 \\
    &\iff 2^{|X|+|Y|+|Z|+|W|} \leq m+|Z|+|W|+|Y| \\
    &\iff 2^{|X|+|Y|+|Z|+|W|} \leq m+|Z|+|W|+\min\{|X|,|Y|\}.
\end{align*}
\qed
\end{proof}
\begin{corollary}
    Suppose $D=uv\Delta_{W}$ as in Theorem 1 in \cite{shi2022few-weight}, then $\mathcal{C}_{D}^{(2)}$ is a one-weight binary linear code with length $2^{|W|}$, dimension $|W|$ and minimum distance $2^{|W|-1}$. Moreover, $\mathcal{C}_{D}^{(2)}$ is distance-optimal if $|W| \geq 2$.
\end{corollary}
\begin{corollary}
    Suppose $D=uv\Delta_{W}^{c}$ as in Theorem 2 in \cite{shi2022few-weight}, then $\mathcal{C}_{D}^{(2)}$ is a two-weight binary linear code with length $2^{m}-2^{|W|}$, dimension $m$ and minimum distance $2^{m-1}-2^{|W|-1}$. Moreover, $\mathcal{C}_{D}^{(2)}$ attains the Griesmer bound so that it is distance-optimal.
\end{corollary}
\begin{corollary}
    Suppose $D=u\mathbb{F}_2^m+v\Delta_L+uv\Delta_N^c$ as in Theorem 5 in \cite{shi2022few-weight}, then $\mathcal{C}_{D}^{(2)}$ is a two-weight binary linear code with length $(2^{m}-2^{|W|})2^{m+|Y|}$, dimension $2m+|Y|$ and minimum distance $(2^{m}-2^{|W|})2^{m+|Y|-1}$. Moreover, $\mathcal{C}_{D}^{(2)}$ attains the Griesmer bound so that it is distance-optimal.
\end{corollary}
\begin{remark}
    The subfield codes of $\mathcal{C}_D$-codes considered in this article have better parameters as compared to the Gray images of $\mathcal{C}_D$-codes studied in (\cite{shi2022few-weight}, Proposition $1$).
\end{remark}
\begin{theorem}\label{minimal_selforthogonal}
    Let $m$ be a positive integer and $X, Y, Z, W \subseteq [m]$. Then Table \ref{summary_table} presents the sufficient conditions for $\mathcal{C}_{D}^{(2)}$ to be self-orthogonal and minimal.
\end{theorem}
\begin{proof}
    The self-orthogonality conditions follow directly from Theorem \ref{self_orthogonality_theorem}. For the minimality condition, we provide proofs for parts $(3)$ and $(4)$ of Theorem \ref{maintheorem}.
    
    Part-$(3)$ Without loss of generality, assume that $\max\{|X|,|Y|\} = |X|$. Then  $\wt_{\max} = (2^{m}-2^{|Y|})2^{m+|Z|+|W|-1}$ and $\wt_{\min}=(2^{m}-2^{|X|}-2^{|Y|})2^{m+|Z|+|W|-1}.$ 
    \begin{equation*}
        \frac{\wt_{\min}}{\wt_{\max}} > \frac{1}{2} \iff 2^{m} > 2^{|Y|}+2^{|X|+1} \impliedby 2^{m} > 3.2^{|X|} \impliedby 2^{m} \geq 2^{|X|+2} \iff |X| \leq m-2.
    \end{equation*}
Part-$(4)$ If $s_{1}< s_{2} < s_{3}$ and $\{s_{1},s_{2},s_{3}\} = \{|X|,|Y|,|Z|\}$, then $\wt_{\min} = (2^{m}-2^{s_{1}})(2^{m}-2^{s_{2}}-2^{s_{3}})2^{m+|W|-1}$ and $\wt_{\max} = (2^{m}-2^{s_{1}})(2^{m}-2^{s_2})2^{m+|W|-1}$. 
\begin{equation*}
        \frac{\wt_{\min}}{\wt_{\max}} > \frac{1}{2} \iff 2^{m} > 2^{s_{2}}+2^{s_{3}+1} \impliedby 2^{m} > 3.2^{s_{3}} \impliedby 2^{m} \geq 2^{s_{3}+2} \iff s_{3} \leq m-2. 
    \end{equation*}
    \qed
    \end{proof}
We now present a few examples to illustrate our results using MAGMA \cite{bosma1997magma}.
\begin{example}
    Take $m=4, X=Y=Z=\emptyset$ and $W= \{1,2,3\}.$ Then $\Delta_K=\Delta_L=\Delta_M=\{(0, 0, 0, 0)\}$ and $\Delta_N=\{(i, j, k, 0): i, j, k \in \mathbb{F}_2\}.$ If $D$ is as in Theorem \ref{maintheorem} $(2)$, then $\mathcal{C}_{D}^{(2)}$ is a two-weight optimal binary linear code having parameters $[120,7,60].$ Moreover, Theorem \ref{minimal_selforthogonal} implies that $\mathcal{C}_{D}^{(2)}$ is minimal and self-orthogonal. The Hamming weight enumerator of $\mathcal{C}_{D}^{(2)}$ is $x^{120}+15x^{56}y^{64}+112x^{60}y^{60}.$ By utilizing the well-known CSS construction \cite{CSS1998Quantum}, one gets the quantum error-correcting code having parameters $[[120,106,3]]$, which is almost-optimal. 
\end{example}
\begin{example}
    Take $m=3, X=Z=W=\emptyset$ and $Y= \{1\}.$ If $D$ is as in Theorem \ref{maintheorem} $(3)$, then $\mathcal{C}_{D}^{(2)}$ is a four-weight optimal binary linear code having parameters $[42,6,20].$ Moreover, Theorem \ref{minimal_selforthogonal} implies that $\mathcal{C}_{D}^{(2)}$ is minimal. The Hamming weight enumerator of $\mathcal{C}_{D}^{(2)}$ is $x^{42}+21x^{22}y^{20}+32x^{21}y^{21}+7x^{18}y^{24}+3x^{14}y^{28}.$
\end{example}
\begin{example}
    Take $m=4, X=Z=\{1\}, Y=\emptyset, W=[m].$ If $D$ is as in Theorem \ref{maintheorem} $(2)$, then $\mathcal{C}_{D}^{(2)}$ is a two-weight optimal binary linear code having parameters $[448,9,224].$ Moreover, Theorem \ref{minimal_selforthogonal} implies that $\mathcal{C}_{D}^{(2)}$ is minimal and self-orthogonal. The Hamming weight enumerator of $\mathcal{C}_{D}^{(2)}$ is $x^{448}+504x^{224}y^{224}+7x^{192}y^{256}.$ By utilizing the well-known CSS construction, one gets the quantum error-correcting code having parameters $[[448,430,3]]$, which is almost-optimal. 
\end{example}
\begin{example}
    For $m\ge 2,$ if $|X|\leq m-2,$ then the code $\mathcal{C}_{D}^{(2)}$ in Theorem \ref{maintheorem} $(2)$ is distance-optimal, minimal and self-orthogonal for any non-empty subsets $Y, Z, W$ of $[m].$ 
\end{example}

\begin{theorem}\label{SRGtheorem}
    Consider the subfield code $\mathcal{C}_{D}^{(2)}$ as in Theorem \ref{maintheorem} (2). Then there exist two families of strongly regular graph, having parameters $$\left( 2^{m+|Y|+|Z|+|W|}, (2^{m}-2^{|X|})2^{|Y|+|Z|+|W|},(2^{m}-2^{|X|+1})2^{|Y|+|Z|+|W|},(2^{m}-2^{|X|})2^{|Y|+|Z|+|W|}\right)$$ and $$\left( 2^{m+|Y|+|Z|+|W|},2^{|X|+|Y|+|Z|+|W|}-1,2^{|X|+|Y|+|Z|+|W|}-2,0\right).$$
\end{theorem}
\begin{proof}
    Here $n=(2^{m}-2^{|X|})2^{|Y|+|Z|+|W|}, k = m+|Y|+|Z|+|W|, w_{1} = 2^{m+|Y|+|Z|+|W|-1}$ and $w_{2} = (2^{m}-2^{|X|})2^{|Y|+|Z|+|W|-1}.$ It is not difficult to show that $d({\mathcal{C}_{D}^{(2)}}^{\perp}) \geq 3$ using the MacWilliams Identities. Invoking Lemma \ref{SRGlemma}, we deduce that the graph corresponding to $\mathcal{C}_{D}^{(2)}$ is strongly regular with parameters
    $$\left( 2^{m+|Y|+|Z|+|W|}, (2^{m}-2^{|X|})2^{|Y|+|Z|+|W|},(2^{m}-2^{|X|+1})2^{|Y|+|Z|+|W|},(2^{m}-2^{|X|})2^{|Y|+|Z|+|W|}\right).$$
    Moreover, its complement graph is also strongly regular with parameters 
    $$\left( 2^{m+|Y|+|Z|+|W|},2^{|X|+|Y|+|Z|+|W|}-1,2^{|X|+|Y|+|Z|+|W|}-2,0\right).$$
    \qed
\end{proof}

\begin{theorem}
    Consider the the subfield code $\mathcal{C}_{D}^{(2)}$ as in Theorem \ref{maintheorem} (6). Then there exist two families of strongly regular graph, having parameters $$\left( 2^{4m}, 2^{4m}-2^{|X|+|Y|+|Z|+|W|},2^{4m}-2^{|X|+|Y|+|Z|+|W|+1},2^{4m}-2^{|X|+|Y|+|Z|+|W|}\right)$$ and $$\left( 2^{4m}, 2^{|X|+|Y|+|Z|+|W|}-1,2^{|X|+|Y|+|Z|+|W|}-2,0\right).$$
\end{theorem}
\begin{proof}
    The proof follows verbatim from Theorem \ref{SRGtheorem}. \qed
\end{proof}

\begin{table}[H]
\centering
\begin{tabular}{c|c}
\hline
Weight & Frequency \\
\hline
$0$ & $1$ \\
\hline
$2^{|X|+|Y|+|Z|+|W|-1} $ & $2^{|X|+|Y|+|Z|+|W|}-1$ \\
\hline
\end{tabular}
\caption{Weight Distribution of Theorem \ref{maintheorem} (1)}
\label{table1}
\end{table}

\begin{table}[H]
\centering
\begin{tabular}{c|c}
\hline
Weight & Frequency \\
\hline
$0$ & $1$ \\
\hline
$2^{m+|Y|+|Z|+|W|-1}$ & $2^{m-|X|}-1$ \\
\hline
$(2^{m}-2^{|X|})2^{|Y|+|Z|+|W|-1} $ & $2^{m+|Y|+|Z|+|W|}-2^{m-|X|} $ \\
\hline

\end{tabular}
\caption{Weight Distribution of Theorem \ref{maintheorem} (2)}
\label{table2}
\end{table}




\begin{table}[H]
\centering
\begin{tabular}{c|c}
\hline
Weight & Frequency \\
\hline
$0$ & $1$ \\
\hline
$(2^{m}-2^{|X|})2^{m+|Z|+|W|-1} $ & $2^{m-|Y|}-1 $ \\
\hline
$(2^{m}-2^{|Y|})2^{m+|Z|+|W|-1} $ & $2^{m-|X|}-1 $ \\
\hline
$(2^{m}-2^{|X|}-2^{|Y|})2^{m+|Z|+|W|-1} $ & $2^{2m-|X|-|Y|}-2^{m-|Y|}-2^{m-|X|}+1 $ \\
\hline
$(2^{m}-2^{|X|})(2^{m}-2^{|Y|})2^{|Z|+|W|-1} $ & $2^{2m+|Z|+|W|}-2^{2m-|X|-|Y|} $ \\
\hline
\end{tabular}
\caption{Weight Distribution of Theorem \ref{maintheorem} (3)}
\label{table6}
\end{table}

\begin{table}[H]
\centering
\begin{tabular}{p{7cm}|p{7.5cm}}
\hline
Weight & Frequency \\
\hline
$0$ & $1$ \\
\hline
$(2^{m}-2^{|X|})(2^{m}-2^{|Y|})2^{m+|W|-1} $ & $2^{m-|Z|}-1 $ \\
\hline
$(2^{m}-2^{|X|})(2^{m}-2^{|Z|})2^{m+|W|-1} $ & $2^{m-|Y|}-1 $ \\
\hline
$(2^{m}-2^{|Y|})(2^{m}-2^{|Z|})2^{m+|W|-1} $ & $2^{m-|X|}-1 $ \\
\hline
$(2^{m}-2^{|X|})(2^{m}-2^{|Y|}-2^{|Z|})2^{m+|W|-1} $ & $2^{2m-|Y|-|Z|}-2^{m-|Y|}-2^{m-|Z|}+1 $ \\
\hline
$(2^{m}-2^{|Y|})(2^{m}-2^{|X|}-2^{|Z|})2^{m+|W|-1} $ & $2^{2m-|X|-|Z|}-2^{m-|X|}-2^{m-|Z|}+1 $ \\
\hline
$(2^{m}-2^{|Z|})(2^{m}-2^{|X|}-2^{|Y|})2^{m+|W|-1} $ & $2^{2m-|X|-|Y|}-2^{m-|X|}-2^{m-|Y|}+1 $ \\
\hline
$(2^{2m}-2^{m+|X|}-2^{m+|Y|}-2^{m+|Z|}+2^{|X|+|Y|}+2^{|Y|+|Z|}+2^{|X|+|Z|})2^{m+|W|-1} $ & $2^{3m-|X|-|Y|-|Z|}-2^{2m-|Y|-|Z|}-2^{2m-|X|-|Z|}-2^{2m-|X|-|Y|}+2^{m-|Z|}+2^{m-|Y|}+2^{m-|X|}-1 $ \\
\hline
$(2^{m}-2^{|X|})(2^{m}-2^{|Y|})(2^{m}-2^{|Z|})2^{|W|-1}$ & $2^{3m+|W|}-2^{3m-|X|-|Y|-|Z|}$ \\
\hline
\end{tabular}
\caption{Weight Distribution of Theorem \ref{maintheorem} (4)}
\label{table12}
\end{table}

\begin{table}[H]
\centering
\scalebox{0.9}{
\begin{tabular}{p{7.5cm}|p{7.5cm}}
\hline
Weight & Frequency \\
\hline
$0$ & $1$ \\
\hline
$(2^{m}-2^{|X|})(2^{m}-2^{|Z|})(2^{m}-2^{|W|})2^{m-1} $ & $2^{m-|Y|}-1 $ \\
\hline
$(2^{m}-2^{|X|})(2^{m}-2^{|Y|})(2^{m}-2^{|W|})2^{m-1} $ & $2^{m-|Z|}-1 $ \\
\hline
$(2^{m}-2^{|X|})(2^{m}-2^{|Y|})(2^{m}-2^{|Z|})2^{m-1} $ & $2^{m-|W|}-1 $ \\
\hline
$(2^{m}-2^{|Y|})(2^{m}-2^{|Z|})(2^{m}-2^{|W|})2^{m-1} $ & $2^{m-|X|}-1 $ \\
\hline
$(2^{m}-2^{|X|})(2^{m}-2^{|W|})(2^{m}-2^{|Y|}-2^{|Z|})2^{m-1} $ & $2^{2m-|Y|-|Z|}-2^{m-|Y|}-2^{m-|Z|}+1 $ \\
\hline
$(2^{m}-2^{|X|})(2^{m}-2^{|Z|})(2^{m}-2^{|Y|}-2^{|W|})2^{m-1} $ & $2^{2m-|Y|-|W|}-2^{m-|Y|}-2^{m-|W|}+1 $ \\
\hline
$(2^{m}-2^{|X|})(2^{m}-2^{|Y|})(2^{m}-2^{|Z|}-2^{|W|})2^{m-1} $ & $2^{2m-|Z|-|W|}-2^{m-|Z|}-2^{m-|W|}+1 $ \\
\hline
$(2^{m}-2^{|Z|})(2^{m}-2^{|W|})(2^{m}-2^{|X|}-2^{|Y|})2^{m-1} $ & $2^{2m-|X|-|Y|}-2^{m-|X|}-2^{m-|Y|}+1 $ \\
\hline
$(2^{m}-2^{|Y|})(2^{m}-2^{|W|})(2^{m}-2^{|X|}-2^{|Z|})2^{m-1} $ & $2^{2m-|X|-|Z|}-2^{m-|X|}-2^{m-|Z|}+1 $ \\
\hline
$(2^{m}-2^{|Y|})(2^{m}-2^{|Z|})(2^{m}-2^{|X|}-2^{|W|})2^{m-1} $ & $2^{2m-|X|-|W|}-2^{m-|X|}-2^{m-|W|}+1 $ \\
\hline
$(2^{m}-2^{|X|})(2^{2m}-2^{m+|Z|}-2^{m+|W|}-2^{m+|Y|}+2^{|Z|+|W|}+2^{|Y|+|Z|}+2^{|Y|+|W|})2^{m-1} $ & $2^{3m-|Y|-|Z|-|W|}-2^{2m-|Y|-|W|}-2^{2m-|Y|-|Z|}-2^{2m-|Z|-|W|}+2^{m-|Y|}+2^{m-|W|}+2^{m-|Z|}-1 $ \\
\hline
$(2^{m}-2^{|W|})(2^{2m}-2^{m+|X|}-2^{m+|Y|}-2^{m+|Z|}+2^{|X|+|Y|}+2^{|Y|+|Z|}+2^{|X|+|Z|})2^{m-1} $ & $2^{3m-|X|-|Y|-|Z|}-2^{2m-|X|-|Y|}-2^{2m-|Y|-|Z|}-2^{2m-|X|-|Z|}+2^{m-|X|}+2^{m-|Y|}+2^{m-|Z|}-1 $ \\
\hline
$(2^{m}-2^{|Z|})(2^{2m}-2^{m+|X|}-2^{m+|Y|}-2^{m+|W|}+2^{|X|+|Y|}+2^{|X|+|W|}+2^{|Y|+|W|})2^{m-1} $ & $2^{3m-|X|-|Y|-|W|}-2^{2m-|X|-|Y|}-2^{2m-|Y|-|W|}-2^{2m-|X|-|W|}+2^{m-|X|}+2^{m-|Y|}+2^{m-|W|}-1 $ \\
\hline
$(2^{m}-2^{|Y|})(2^{2m}-2^{m+|X|}-2^{m+|Z|}-2^{m+|W|}+2^{|X|+|Z|}+2^{|X|+|W|}+2^{|Z|+|W|})2^{m-1} $ & $2^{3m-|X|-|Z|-|W|}-2^{2m-|X|-|Z|}-2^{2m-|Z|-|W|}-2^{2m-|X|-|W|}+2^{m-|X|}+2^{m-|Z|}+2^{m-|W|}-1 $ \\
\hline
$(2^{3m}-2^{2m+|Z|}-2^{2m+|W|}-2^{2m+|X|}-2^{2m+|Y|}+2^{m+|Z|+|W|}+2^{m+|X|+|Z|}+2^{m+|X|+|W|}+2^{m+|Z|+|Y|}+2^{m+|Y|+|W|}+2^{m+|X|+|Y|}-2^{|X|+|Z|+|W|}-2^{|Y|+|Z|+|W|}-2^{|X|+|Y|+|Z|}-2^{|X|+|Y|+|W|})2^{m-1} $ & $2^{4m-|X|-|Y|-|Z|-|W|}-2^{3m-|X|-|Y|-|Z|}-2^{3m-|Y|-|Z|-|W|}-2^{3m-|X|-|Z|-|W|}-2^{3m-|X|-|Y|-|W|}+2^{2m-|Y|-|Z|}+2^{2m-|X|-|Z|}+2^{2m-|Z|-|W|}+2^{2m-|X|-|Y|}+2^{2m-|Y|-|W|}+2^{2m-|X|-|W|}-2^{m-|X|}-2^{m-|Y|}-2^{m-|Z|}-2^{m-|W|}+1 $ \\
\hline
$(2^{m}-2^{|X|})(2^{m}-2^{|Y|})(2^{m}-2^{|Z|})(2^{m}-2^{|W|})2^{-1}$ & $2^{4m}-2^{4m-|X|-|Y|-|Z|-|W|}$ \\
\hline
\end{tabular}}
\caption{Weight Distribution of Theorem \ref{maintheorem} (5)}
\label{table16}
\end{table}

\begin{table}[H]
\centering
\begin{tabular}{c|c}
\hline
Weight & Frequency \\
\hline
$0$ & $1$ \\
\hline
$2^{4m-1}$ & $2^{4m-|X|-|Y|-|Z|-|W|}-1$ \\
\hline
$2^{4m-1}-2^{|X|+|Y|+|Z|+|W|-1}$ & $2^{4m}-2^{4m-|X|-|Y|-|Z|-|W|} $ \\
\hline
\end{tabular}
\caption{Weight Distribution of Theorem \ref{maintheorem} (6)}
\label{table17}
\end{table} 

\begin{landscape}
    \begin{table}
     \centering
     \scalebox{0.50}{
    \begin{tabular}{|c|c|c|c|c|c|}
    \hline
    Defining Set & Parameters & Weight & Distance-Optimal & Minimal & Self-Orthogonal \\
    \hline
$D=\mathbf{b}_{1}\Delta_{X}+\mathbf{b}_{2}\Delta_{Y}+\mathbf{b}_{3}\Delta_{Z}+\mathbf{b}_{4}\Delta_{W}$ & $[2^{|X|+|Y|+|Z|+|W|,|X|+|Y|+|Z|+|W|,2^{|X|+|Y|+|Z|+|W|}-1]}$ & $1$ & Yes, if $|X|+|Y|+|Z|+|W| \geq 2$ & Yes & Yes, if $|X|+|Y|+|Z|+|W| \geq 3$ \\
    \hline
$D=\mathbf{b}_{1}\Delta_{X}^{c}+\mathbf{b}_{2}\Delta_{Y}+\mathbf{b}_{3}\Delta_{Z}+\mathbf{b}_{4}\Delta_{W}$ & $[(2^m-2^{|X|})2^{|Y|+|Z|+|W|},m+|Y|+|Z|+|W|,(2^m-2^{|X|})2^{|Y|+|Z|+|W|-1}]$ & $2$ & Yes & Yes, if $|X|\leq m-2$ & Yes, if $|Y|+|Z|+|W| \geq 3$ \\
    \hline
$D=\mathbf{b}_{1}\Delta_{X}+\mathbf{b}_{2}\Delta_{Y}^{c}+\mathbf{b}_{3}\Delta_{Z}+\mathbf{b}_{4}\Delta_{W}$ & $[(2^m-2^{|Y|})2^{|X|+|Z|+|W|},m+|X|+|Z|+|W|,(2^m-2^{|Y|})2^{|X|+|Z|+|W|-1}]$ & $2$ & Yes & Yes, if $|Y|\leq m-2$ & Yes, if $|X|+|Z|+|W| \geq 3$ \\
    \hline $D=\mathbf{b}_{1}\Delta_{X}+\mathbf{b}_{2}\Delta_{Y}+\mathbf{b}_{3}\Delta_{Z}^{c}+\mathbf{b}_{4}\Delta_{W}$ & $[(2^m-2^{|Z|})2^{|X|+|Y|+|W|},m+|X|+|Y|+|W|,(2^m-2^{|Z|})2^{|X|+|Y|+|W|-1}]$ & $2$ & Yes & Yes, if $|Z|\leq m-2$ & Yes, if $|X|+|Y|+|W| \geq 3$ \\
    \hline
    $D=\mathbf{b}_{1}\Delta_{X}+\mathbf{b}_{2}\Delta_{Y}+\mathbf{b}_{3}\Delta_{Z}+\mathbf{b}_{4}\Delta_{W}^{c}$ & $[(2^m-2^{|W|})2^{|X|+|Y|+|Z|},m+|X|+|Y|+|Z|,(2^m-2^{|W|})2^{|X|+|Y|+|Z|-1}]$ & $2$ & Yes & Yes, if $|W|\leq m-2$ & Yes, if $|X|+|Y|+|Z| \geq 3$ \\
    \hline
    $D=\mathbf{b}_{1}\Delta_{X}^{c}+\mathbf{b}_{2}\Delta_{Y}^{c}+\mathbf{b}_{3}\Delta_{Z}+\mathbf{b}_{4}\Delta_{W}$ & $[(2^m-2^{|X|})(2^m-2^{|Y|})2^{|Z|+|W|},2m+|Z|+|W|,(2^m-2^{|X|}-2^{|Y|})2^{m+|Z|+|W|-1}]$ & $4$ & Yes, if $2^{|X|+|Y|+|Z|+|W|} \leq m+|Z|+|W|+\min\{|X|,|Y|\}$ & Yes, if $\max\{|X|,|Y|\}\leq m-2$ & Yes, if $|Z|+|W| \geq 3$ \\
    \hline
    $D=\mathbf{b}_{1}\Delta_{X}^{c}+\mathbf{b}_{2}\Delta_{Y}+\mathbf{b}_{3}\Delta_{Z}^{c}+\mathbf{b}_{4}\Delta_{W}$ & $[(2^m-2^{|X|})(2^m-2^{|Z|})2^{|Y|+|W|},2m+|Y|+|W|,(2^m-2^{|X|}-2^{|Z|})2^{m+|Y|+|W|-1}]$ & $4$ & Yes, if $2^{|X|+|Y|+|Z|+|W|} \leq m+|Y|+|W|+\min\{|X|,|Z|\}$ & Yes, if $\max\{|X|,|Z|\}\leq m-2$ & Yes, if $|Y|+|W| \geq 3$ \\
    \hline
    $D=\mathbf{b}_{1}\Delta_{X}^{c}+\mathbf{b}_{2}\Delta_{Y}+\mathbf{b}_{3}\Delta_{Z}+\mathbf{b}_{4}\Delta_{W}^{c}$ & $[(2^m-2^{|X|})(2^m-2^{|W|})2^{|Y|+|Z|},2m+|Y|+|Z|,(2^m-2^{|X|}-2^{|W|})2^{m+|Y|+|Z|-1}]$ & $4$ & Yes, if $2^{|X|+|Y|+|Z|+|W|} \leq m+|Y|+|Z|+\min\{|X|,|W|\}$ & Yes, if $\max\{|X|,|W|\}\leq m-2$ & Yes, if $|Y|+|Z| \geq 3$ \\
    \hline
    $D=\mathbf{b}_{1}\Delta_{X}+\mathbf{b}_{2}\Delta_{Y}^{c}+\mathbf{b}_{3}\Delta_{Z}^{c}+\mathbf{b}_{4}\Delta_{W}$ & $[(2^m-2^{|Y|})(2^m-2^{|Z|})2^{|X|+|W|},2m+|X|+|W|,(2^m-2^{|Y|}-2^{|Z|})2^{m+|X|+|W|-1}]$ & $4$ & Yes, if $2^{|X|+|Y|+|Z|+|W|} \leq m+|X|+|W|+\min\{|Y|,|Z|\}$ & Yes, if $\max\{|Y|,|Z|\}\leq m-2$ & Yes, if $|X|+|W| \geq 3$ \\
    \hline
    $D=\mathbf{b}_{1}\Delta_{X}+\mathbf{b}_{2}\Delta_{Y}^{c}+\mathbf{b}_{3}\Delta_{Z}+\mathbf{b}_{4}\Delta_{W}^{c}$ & $[(2^m-2^{|Y|})(2^m-2^{|W|})2^{|X|+|Z|},2m+|X|+|Z|,(2^m-2^{|Y|}-2^{|W|})2^{m+|X|+|Z|-1}]$ & $4$ & Yes, if $2^{|X|+|Y|+|Z|+|W|} \leq m+|X|+|Z|+\min\{|Y|,|W|\}$ & Yes, if $\max\{|Y|,|W|\}\leq m-2$ & Yes, if $|X|+|Z| \geq 3$ \\
    \hline
    $D=\mathbf{b}_{1}\Delta_{X}+\mathbf{b}_{2}\Delta_{Y}+\mathbf{b}_{3}\Delta_{Z}^{c}+\mathbf{b}_{4}\Delta_{W}^{c}$ & $[(2^m-2^{|Z|})(2^m-2^{|W|})2^{|X|+|Y|},2m+|X|+|Y|,(2^m-2^{|Z|}-2^{|W|})2^{m+|X|+|Y|-1}]$ & $4$ & Yes, if $2^{|X|+|Y|+|Z|+|W|} \leq m+|X|+|Y|+\min\{|Z|,|W|\}$ & Yes, if $\max\{|Z|,|W|\}\leq m-2$ & Yes, if $|X|+|Y| \geq 3$ \\
    \hline
    $D=\mathbf{b}_{1}\Delta_{X}^{c}+\mathbf{b}_{2}\Delta_{Y}^{c}+\mathbf{b}_{3}\Delta_{Z}^{c}+\mathbf{b}_{4}\Delta_{W}$ & $[(2^{m}-2^{|X|})(2^m-2^{|Y|})(2^m-2^{|Z|})2^{|W|},3m+|W|]$ & $8$ & - & Yes, if $\max\{|X|,|Y|,|Z|\}\leq m-2$ & Yes, if $|W| \geq 3$ \\
    \hline
    $D=\mathbf{b}_{1}\Delta_{X}^{c}+\mathbf{b}_{2}\Delta_{Y}^{c}+\mathbf{b}_{3}\Delta_{Z}+\mathbf{b}_{4}\Delta_{W}^{c}$ & $[(2^{m}-2^{|X|})(2^m-2^{|Y|})(2^m-2^{|W|})2^{|Z|},3m+|Z|]$ & $8$ & - & Yes, if $\max\{|X|,|Y|,|W|\}\leq m-2$ & Yes, if $|Z| \geq 3$ \\
    \hline
    $D=\mathbf{b}_{1}\Delta_{X}^{c}+\mathbf{b}_{2}\Delta_{Y}+\mathbf{b}_{3}\Delta_{Z}^{c}+\mathbf{b}_{4}\Delta_{W}^{c}$ & $[(2^{m}-2^{|X|})(2^m-2^{|Z|})(2^m-2^{|W|})2^{|Y|},3m+|Y|]$ & $8$ & - & Yes, if $\max\{|X|,|Z|,|W|\}\leq m-2$ & Yes, if $|Y| \geq 3$ \\
    \hline
    $D=\mathbf{b}_{1}\Delta_{X}+\mathbf{b}_{2}\Delta_{Y}^{c}+\mathbf{b}_{3}\Delta_{Z}^{c}+\mathbf{b}_{4}\Delta_{W}^{c}$ & $[(2^{m}-2^{|Y|})(2^m-2^{|Z|})(2^m-2^{|W|})2^{|X|},3m+|X|]$ & $8$ & - & Yes, if $\max\{|Y|,|Z|,|W|\}\leq m-2$ & Yes, if $|X| \geq 3$ \\
    \hline
    $D=\mathbf{b}_{1}\Delta_{X}^{c}+\mathbf{b}_{2}\Delta_{Y}^{c}+\mathbf{b}_{3}\Delta_{Z}^{c}+\mathbf{b}_{4}\Delta_{W}^{c}$ & $[(2^{m}-2^{|X|})(2^m-2^{|Y|})(2^m-2^{|Z|})(2^{m}-2^{|W|}),4m]$ & $16$ & - & Yes, if $\max\{|X|,|Y|,|Z|,|W|\}\leq m-2$ & Yes, if $X,Y,Z,W\neq \emptyset$  \\
    \hline
    $D=(\mathbf{b}_{1}\Delta_{X}+\mathbf{b}_{2}\Delta_{Y}+\mathbf{b}_{3}\Delta_{Z}+\mathbf{b}_{4}\Delta_{W})^{c}$ & $[2^{4m}-2^{|X|+|Y|+|Z|+|W|},4m,2^{4m}-2^{|X|+|Y|+|Z|+|W|-1}]$ & $2$ & Yes & Yes, if $|X|+|Y|+|Z|+|W| \leq 4m-2$ & Yes, if $|X|+|Y|+|Z|+|W| \geq 3$ \\
    \hline
    \end{tabular}}
    \caption{Binary linear codes from simplicial complexes}
    \label{summary_table}
    \end{table}
    \end{landscape}

\section{Conclusion}\label{Section 5}
 In this article, we considered $\mathcal{C}_D$-codes over $\mathcal{R}:=\mathbb{F}_{2}[x,y]/\langle x^2, y^2, xy-yx\rangle$ and we employ a trace map from $\mathcal{R} $ to $\mathbb{F}_2$ to investigate their binary subfield codes, where the defining set $D$ is constructed from a simplicial complex. This article contributes to the literature by determining the Hamming weight distribution of these codes. We also get certain infinite families of distance-optimal codes. In addition, we deduce sufficient conditions for them to be minimal and self-orthogonal. Table \ref{summary_table} summarizes these results.

In the future, one may investigate subfield codes of $\mathcal{C}_D$-codes over different alphabets. We considered only simplicial complexes having a single maximal element, as extending our method to complexes with two maximal elements leads to tedious computations. Defining sets based on simplicial complexes with two maximal elements over $\mathcal{R}$ remains an interesting avenue for future exploration.

\section*{Acknowledgements}
The first author expresses gratitude to MHRD, India, for financial support in the form of a Junior Research Fellowship at the Indian Institute of Technology, Delhi. The third author acknowledges the support of the Council of Scientific and Industrial Research (CSIR) India, under grant no. 09/0086(13310)/2022-EMR-I.
\section*{Declarations}
\subsection*{Conflict of Interest}
All authors declare that they have no conflict of interest.

\bibliographystyle{abbrv}
\bibliography{references}
    
\end{document}